\documentclass[12pt]{elsarticle}

\usepackage{definitions}
\usepackage{heppennames2}
\usepackage{sidecap}
\usepackage[algo2e]{algorithm2e}
\usepackage{algorithmic}
\usepackage{algorithm}
\usepackage{float}
\newfloat{algorithm}{t}{lop}
\usepackage{bbold}
\usepackage[bb=ams]{mathalfa}
\usepackage{dsfont}
\usepackage{amsfonts}
\usepackage{bm}
\usepackage{hyperref}
\usepackage{multirow}
\usepackage{amsmath}
\usepackage{setspace}
\usepackage{listings}
\usepackage{pythonhighlight}
\usepackage{graphicx}
\DeclareMathOperator*{\argmax}{arg\,max}

\let\OLDthebibliography\thebibliography

\renewcommand\thebibliography[1]{
  \OLDthebibliography{#1}
  \setlength{\parskip}{0pt}
  \setlength{\itemsep}{0pt plus 0.3ex}
}

\journal{Computer Physics Communications}
\begin{document}

\begin{frontmatter}
\title{Tree boosting for learning EFT parameters}

\author[a]{Suman Chatterjee}
\author[b]{Nikolaus Frohner}
\author[a]{Lukas Lechner}
\author[a]{Robert Sch\"ofbeck}
\author[a]{Dennis Schwarz}

\cortext[x]{\textit{E-mail addresses:} suman.chatterjee@oeaw.ac.at, nikolaus.frohner@tuwien.ac.at, lukas.lechner@oeaw.ac.at, robert.schoefbeck@oeaw.ac.at, dennis.schwarz@oeaw.ac.at}

\address[a]{Institute of High Energy Physics (HEPHY), Austrian Academy of Sciences (\"{O}AW), Nikolsdorfer Gasse 18, 1050 Vienna, Austria}
\address[b]{TU Wien, Karlsplatz 13, 1040 Vienna, Austria}

\date{\today}

\begin{abstract}
We present a new tree boosting algorithm designed for the measurement of parameters in the context of effective field theory~(EFT). 
To construct the algorithm, we interpret the optimized loss function of a traditional decision tree as the maximal Fisher information in Poisson counting experiments.
We promote the interpretation to general EFT predictions and develop a suitable boosting method.
The resulting ``Boosted Information Tree'' algorithm approximates the score, the derivative of the log-likelihood function with respect to the parameter.
It thus provides a sufficient statistic in the vicinity of a reference point in parameter space where the estimator is trained.
The training exploits per-event information of likelihood ratios for different theory parameter values available in the simulated EFT data sets.

\end{abstract}


\end{frontmatter}
\newpage
{\bf PROGRAM SUMMARY}

\begin{small}
\noindent
{\em Program Title: BIT (Boosted Information Trees)}                                          \\
{\em CPC Library link to program files:} (to be added by Technical Editor) \\
{\em Developer's repository link: \url{https://github.com/HephyAnalysisSW/BIT}}  \\
{\em Code Ocean capsule:} (to be added by Technical Editor)\\
{\em Licensing provisions:} GPLv3\\
{\em Programming language: Python2 and Python3}\\
{\em Nature of problem: Providing a discriminator for parameter estimation in the context of the standard model effective field theory.}\\
{\em Solution method: A tree-based algorithm exploits ``augmented'' information of the simulated training data set to regress in the score function and thereby provides a sufficient test statistic of an EFT parameter.}\\
\end{small}

\section{Introduction}

An important front at the Large Hadron Collider (LHC) experiments is the search for indirect effects in standard model (SM) measurements induced by phenomena beyond the SM (BSM).
If the BSM energy scale is much higher than the reach of the LHC,  indirect non-resonant effects at lower energies may still be discernible via dispersed and subtle differences in the spectra of kinematic observables.
These are accessible by performing precision measurements of differential cross sections.
The SM effective field theory (SM-EFT)~\cite{Burgess:2007pt,Giudice:2007fh,Grinstein:1991cd,Brivio:2017vri,deFlorian:2016spz,Grzadkowski:2010es} parameterizes such deviations and has become the leading generative model in the first attempts of a global interpretation of collider results~\cite{Falkowski:2019hvp,DeBlas:2019ehy,Ellis:2020unq,Dawson:2020oco,Ethier:2021bye,Ethier:2021ydt}. 
The SM-EFT has 2499 degrees of freedom, among which 59 are flavor diagonal and respect baryon and lepton number conservation~\cite{Grzadkowski:2010es,Buchmuller:1985jz}. 

Particle physics event generation, both, for SM-EFT and in general, consists of consecutive steps involving simulation tools for the hard interaction, the parton shower, the matrix element~(ME) matching, the hadronization of strongly interacting partons, and the detector response to the stable particles.
The authors of Refs.~\cite{Cranmer:2015bka,Brehmer:2018kdj,Brehmer:2018eca,Brehmer:2018hga,Brehmer:2019xox} employ this structure to show how mean squared error~(MSE) loss functions can regress on the SM-EFT likelihood, the likelihood ratio between different SM-EFT hypotheses, or the gradient in SM-EFT parameter space of the likelihood function, i.e., the score vector. 
The key element is additional information from Monte-Carlo simulations that characterizes the dependence of the likelihood function on the SM-EFT parameters. 
This augmented data is used in a number of sophisticated machine learning methods based on neural networks.

Tree boosting~\cite{Friedman:2001wbq, Friedman:2002we,10.5555/3009657.3009730,Mason99boostingalgorithms}, on the other hand, has received much less attention in the context of probing EFT operators.
Boosting algorithms provide a strong learner by iteratively training an ensemble of weak learners to the pseudo-residuals of the previous iteration step.
Traditional Boosted Decision Trees~(BDTs) for classification tasks are trained with discretely categorized data using, e.g., samples of simulated signal and background events and predict the target label. 
The BDTs exhibit robust performance in a wide range of classification problems in high-energy physics~(HEP).
In the context of EFT, tree-based discriminators have not been a focus of dedicated development.
A reason may lie in the absence, so far, of a simple strategy to exploit the augmented data in the training.




In this work, we change that and develop a tree-based algorithm trained on a simulated event sample generated at a single EFT parameter point.
We work with a single scalar parameter, possibly after a linear basis transformation, such that the hypotheses correspond to the two scalar values $\theta_\textrm{0}$ and $\theta$. 
The resulting algorithm is called ``Boosted Information Tree''~(BIT), consisting of a weak learner and a boosting algorithm. 
Its loss function 
uses the augmented data to regresses to the score function. The augmented data are per-event weight functions with a simple analytic dependence on the EFT theory parameters.
By approximating the score function, the BIT discriminator provides a locally sufficient statistic of the parameter.

The structure of the article is as follows. 
In Sec.~\ref{sec:previous-work}, we give an overview of the relevant previous work. A detailed description of the structure of the augmented data and the training  data  set  is  provided in  Sec.~\ref{sec:data}. 
In Sec.~\ref{sec:algo-motivation}, we interpret the fit of a traditional decision tree, the weak learner of a BDT, in terms of an optimization of the Fisher information in Poisson counting experiments. 
This interpretation serves as a starting point for constructing the algorithm in Sec.~\ref{sec:algo-construction}.
The approximation of the score function is validated in analytically tractable toy models, discussed in Sec.~\ref{sec:toys}. 
Conclusions and outlook are provided in Sec.~\ref{sec:conclusion}.
Public code implementing the algorithm is available on GitHub~\cite{BIT-algo}, and a brief introduction to the interface is provided in~\ref{sec:python}. 

\section{Related work}\label{sec:previous-work}
Below we briefly mention the algorithms developed in the recent past and are closely related to the method we report in this article.
\subsection{SALLY/SALLINO}
In a series of publications, the authors of Refs.~\cite{Brehmer:2018kdj,Brehmer:2018eca,Brehmer:2018hga,Brehmer:2019xox} introduce algorithms that use augmented data from the event simulation to regress to the true likelihood or the true score even if the quantities are intractable. 
It is shown how intractable factors in the likelihood, related to the parton shower, the ME matching, the detector simulation, etc., cancel and the augmented data is subsequently leveraged to define a number of sophisticated inference techniques based on neural networks. 
This methodology is also a foundation of the present work. 
At variance with the neural network approximation, we focus on the training of tree-based discriminators. 
The BIT operates in the same setting as the SALLY and SALLINO algorithms.

\subsection{ INFERNO} 
In Ref.~\cite{DeCastro:2018psv}, a non-linear summary statistic is constructed by minimizing inference-motivated losses via stochastic gradient descent. 
Similar to the present work, the algorithm uses the Fisher information of the parameters of interest. 
In contrast to our work, a neural network is trained, and the INFERNO training loss is chosen to minimize the expected variance in the context of parameter regression.

\subsection{ Binary classifier metrics and weight derivative regression} 
In Refs.~\cite{Valassi:2019uhy,Valassi:2020deh}, various measures for classification and regression performance that involve the Fisher information in Poisson counting experiments are introduced. 
The interpretation of a decision tree with the Gini impurity as a Fisher information-optimal cross section measurement is the starting point of the development of the BIT. 
We do not consider the various ratios of the Fisher information at the different stages of an estimation procedure. 
In the present work, we merely follow up on the proposal to generalize node split criteria towards loss functions involving the Fisher information and present a concrete boosted tree algorithm.  

\section{Structure of the training data}\label{sec:data}

At the level of the hard scattering, event simulation of squared MEs provides a parton-level configuration ${\boldsymbol z}\sim p({\boldsymbol z}|{\boldsymbol \theta})$, where the vector ${\boldsymbol \theta}$ denotes the EFT parameters and $p({\boldsymbol z}|{\boldsymbol \theta})$ the parton-level likelihood.
The generic structure of a matrix element in terms of Feynman amplitudes with a single EFT operator insertion leads to predicted differential cross sections of the general form\footnote{EFT operators that mix with the SM kinetic terms can lead, in principle, to non-polynomial dependence on $\boldsymbol{\theta}$. This behavior does not spoil the following discussion, and there are SM-EFT models that conveniently support the truncation of kinetic mixing effects and propagator corrections to a configurable polynomial order, e.g., Ref.~\cite{Brivio:2020onw}. }
\begin{eqnarray}
\textrm{d}\sigma({\boldsymbol\theta})&\propto& |\mathcal{M}_{\textrm{SM}}({\boldsymbol z})+\theta_a\mathcal{M}^a_{\textrm{BSM}}({\boldsymbol z})|^2 \textrm{d}{\boldsymbol z}.\label{eq:poly-xsec}
\end{eqnarray}
The leading BSM contribution is $2\theta_a\textrm{Re}\left(\mathcal{M}^\ast({\boldsymbol z})_{\textrm{SM}} \mathcal{M}^a_{\textrm{BSM}}({\boldsymbol z})\right)$ and describes the interference of the BSM and SM amplitudes. It is the only term in the expansion in $\boldsymbol{\theta}$ that does not also receive contributions from EFT operators with higher mass dimensions.
We write the likelihood $p(\boldsymbol{z}|\boldsymbol{\theta})$, the likelihood ratio $ r(\boldsymbol{z}|\boldsymbol{\theta},\boldsymbol{\theta}_0)$, and the score vector $\boldsymbol{t}(\boldsymbol{z}|\boldsymbol{\theta})$ as
\begin{eqnarray}
    p(\boldsymbol{z}|\boldsymbol{\theta})&=&\frac{1}{\sigma(\boldsymbol{\theta})}\frac{\textrm{d}\sigma( {\boldsymbol\theta})}{\textrm{d}\boldsymbol{z}},\label{eq:likelihood}\\
    r(\boldsymbol{z}|\boldsymbol{\theta},\boldsymbol{\theta}_0)&=&\frac{\textrm{d}\sigma({\boldsymbol\theta})/\textrm{d}\boldsymbol{z}}{\textrm{d}\sigma( {\boldsymbol\theta_\textrm{0}})/\textrm{d}\boldsymbol{z}}\cdot\frac{\sigma(\boldsymbol{\theta}_\textrm{0})}{\sigma(\boldsymbol{\theta})},\;\textrm{and}\label{eq:likelihood-ratio}\\
   \boldsymbol{t}(\boldsymbol{z}|\boldsymbol{\theta})&=&\boldsymbol\nabla_\theta \log  p(\boldsymbol{z}|\boldsymbol{\theta}),\label{eq:score}
\end{eqnarray}
where $\sigma(\boldsymbol{\theta})$ and $\textrm{d}\sigma({\boldsymbol\theta})/\textrm{d}\boldsymbol{z}$ are the total and the differential parton-level cross sections, respectively.

The simple quadratic structure in Eq.~\ref{eq:poly-xsec} has important practical implications:
If a sample of $N$ events $\{\boldsymbol{z}_i\}_{i=1}^N$ is generated from the probability density function~(PDF) in Eq.~\ref{eq:likelihood} at a reference parameter point $\boldsymbol{\theta}_\textrm{ref}$, the value of the matrix element can be reevaluated at linearly independent basis parameter points.
Each simulated event is then augmented with a weight function $w_i(\boldsymbol{\theta})$ that is polynomial in $\boldsymbol{\theta}$ with coefficients determined from Eq.~\ref{eq:poly-xsec}, evaluated at a sufficiently large number of basis parameter points~\cite{Brehmer:2018eca}. 
We can choose an overall normalization~(scaling) of the weight functions to the total number of predicted events for an arbitrarily chosen luminosity $\mathcal{L}$, i.e., $\sum_{i=1}^N w_i(\boldsymbol{\theta})=\mathcal{L}\sigma(\boldsymbol{\theta})$. 
In this ``cross section normalization'', the differential parton-level cross section, integrated over a small phase space volume $\Delta \boldsymbol{z}$, is approximated by
\begin{equation}
    \int_{\Delta \boldsymbol{z}}\frac{\textrm{d}\sigma( {\boldsymbol\theta})}{\textrm{d}\boldsymbol{z}}\textrm{d}\boldsymbol{z}\approx\frac{\textrm{d}\sigma( {\boldsymbol\theta})}{\textrm{d}\boldsymbol{z}}\Delta \boldsymbol{z}\approx\frac{1}{\mathcal{L}}\sum_{z_i\in\Delta \boldsymbol{z}} w_i(\boldsymbol{\theta})\label{eq:diff-xsec-approx}
\end{equation} 
provided that $N$ is sufficiently large and $\textrm{d}\sigma( {\boldsymbol\theta})/\textrm{d}\boldsymbol{z}$ does not vary strongly in the region $\Delta \boldsymbol{z}$. 
The weight functions $w_i(\boldsymbol{\theta})$ encode the $\boldsymbol{\theta}$-dependence of all predictions obtained from simulation.

We can also remove the $\boldsymbol{\theta}$-dependence of the sample's normalization by scaling $\omega_i=w_i N/(\mathcal{L}\sigma(\boldsymbol{\theta}))$, i.e., with a factor common to all events. 
In this ``PDF normalization'', we obtain an approximation of the integral over $\Delta \boldsymbol{z}$  of the PDF as
\begin{equation}
    \int_{\Delta \boldsymbol{z}}p(\boldsymbol{z}|\boldsymbol{\theta})\textrm{d}\boldsymbol{z}\approx\frac{1}{N}\sum_{z_i\in\Delta \boldsymbol{z}} \omega_i(\boldsymbol{\theta}). \label{eq:diff-PDF-approx}
\end{equation} 
Via Eq.~\ref{eq:poly-xsec}--\ref{eq:score}, the weight functions are related to the likelihood ratio and the score function evaluated at the parton-level configuration $\boldsymbol{z}_i$ by
\begin{eqnarray}
r(\boldsymbol{z}_i|\boldsymbol{\theta},\boldsymbol{\theta}_\textrm{ref}) &=& \frac{\omega_i(\boldsymbol{\theta})}{\omega_i(\boldsymbol{\theta}_\textrm{ref})}=\frac{w_i(\boldsymbol{\theta})}{w_i(\boldsymbol{\theta}_\textrm{ref})}\frac{\sigma(\boldsymbol{\theta}_\textrm{ref})}{\sigma(\boldsymbol{\theta})},\label{eq:simu-r}\\
t(\boldsymbol{z}_i|\boldsymbol{\theta}) &=& \boldsymbol\nabla_\theta \log \omega_i(\boldsymbol{\theta})=\boldsymbol\nabla_\theta \log \frac{w_i(\boldsymbol{\theta})}{\sigma(\boldsymbol{\theta})}\label{eq:simu-t}.
\end{eqnarray}

The gain in computational efficiency from beating the curse of dimensionality via the simple $\boldsymbol{\theta}$-dependence in Eqs.~\ref{eq:likelihood}--\ref{eq:simu-t} is enormous. For example, the effects from single insertions of the 15 operators in top quark physics~\cite{Aguilar-Saavedra:2018ksv} require only 136 float numbers per event to compute $\sigma(\boldsymbol{\theta})$, $\textrm{d}\sigma( {\boldsymbol\theta})/\textrm{d}\boldsymbol{z}$, the likelihood, likelihood ratios, and the score vector in the full 15-dimensional space of theory parameters. Assuming that a traditional sampling of the theory parameter space uses a grid of $M$ points in each dimension, a factor of $M^{15}$ is saved. The BIT algorithm will exploit this efficient structure.


Next, we turn our attention to the detector level, i.e. to observables including the simulation of the hard interaction, the parton shower, ME matching, the hadronization of strongly interacting partons, and the detector response to the stable particles.
The event simulation thus provides an event with parton-level configuration $\boldsymbol{z}$ with an observable detector-level feature vector ${\boldsymbol x}$.
The configuration spaces of both, observed event features $\boldsymbol{x}$ and parton-level configuration $\boldsymbol{z}$, are high-dimensional with large numbers of continuous and discrete variables.
Denoting a detector-level selection by $j$, we obtain expected Poisson yields from the simulated sample as $\lambda_j(\boldsymbol{\theta})=\sum_{\boldsymbol{x}_i\in j}w_i(\boldsymbol\theta)$. 
The augmented data set in the cross section normalization (Eq.~\ref{eq:diff-xsec-approx}) is then simply
\begin{equation}
\{w_i( \boldsymbol{\theta}),\mathbf{x}_i,\mathbf{z}_i\}_{i=1}^N.
\end{equation} 
When normalizing to the PDF~(Eq.~\ref{eq:diff-PDF-approx}), we replace $w_i$ by $\omega_i$ to obtain the equivalent data set
\begin{equation}
\{\omega_i( \boldsymbol{\theta}),\mathbf{x}_i,\mathbf{z}_i\}_{i=1}^N.
\end{equation}

A conceptual point pertains to the fact that $\boldsymbol{x}$ and $\boldsymbol{z}$ are jointly available for each simulated event.
Thus, the likelihood $r({\boldsymbol x},{\boldsymbol z}|{\boldsymbol \theta_\textrm{1}},{\boldsymbol \theta}_\textrm{2})$ and the score vector ${\boldsymbol t}({\boldsymbol x},{\boldsymbol z}|{\boldsymbol \theta})=\nabla_{\boldsymbol \theta}\log p({\boldsymbol x},{\boldsymbol z}|{\boldsymbol \theta})$ available from the simulation are jointly conditional on the vectors $\boldsymbol{x}$ and $\boldsymbol{z}$ for arbitrary ${\boldsymbol \theta}$, ${\boldsymbol \theta_\textrm{1}}$, and ${\boldsymbol \theta_\textrm{2}}$. 
According to the Neyman-Pearson lemma, however, it is the true likelihood ratio $r(\boldsymbol{x}|\boldsymbol{\theta}_1,\boldsymbol{\theta}_2)$ that is the optimal test statistic to discriminate between two hypotheses $\boldsymbol{\theta}_1$ and $\boldsymbol{\theta}_2$.
In the case that the parton shower, the ME matching prescription, and the electromagnetic and strong interactions in the detector simulation do not depend on $\boldsymbol{\theta}$,
the true likelihood can be factorized as $p(\boldsymbol{x}|\boldsymbol{\theta})\equiv\int\,p(\boldsymbol{x},\boldsymbol{z})\textrm{d}\boldsymbol{z}=\int\,p(\boldsymbol{x}|\boldsymbol{z})p(\boldsymbol{z}|\boldsymbol{\theta})\textrm{d}\boldsymbol{z}$, but the probability $p(\boldsymbol{x}|\boldsymbol{z})$ to observe a feature vector $\boldsymbol{x}$ given the (latent) parton-level configuration $\boldsymbol{z}$ is intractable. 

So can we still exploit the simple parton-level polynomial structure given only joint quantities while $p(\boldsymbol{x}|\boldsymbol{z})$ is intractable? The authors of Refs.~\cite{Cranmer:2015bka,Brehmer:2018kdj,Brehmer:2018eca,Brehmer:2018hga,Brehmer:2019xox} show how MSE loss functions regress on the true likelihood ratio $r({\boldsymbol x}|{\boldsymbol \theta}_1,{\boldsymbol \theta}_2)$ and the true score ${\boldsymbol t}({\boldsymbol x}|{\boldsymbol \theta})=\boldsymbol{\nabla}_{\boldsymbol \theta}\log p({\boldsymbol x}|{\boldsymbol \theta})$, given only the joint quantities that are also conditional on the parton-level configuration ${\boldsymbol z}$. 
The intractable factors in the likelihood, related to the parton shower, the ME matching, the detector simulation, etc., are shown to cancel~\cite{Brehmer:2018eca} provided that the EFT modifications do not enter $p(\boldsymbol{x}|\boldsymbol{z})$ either directly or indirectly via loop corrections.
We will briefly return to this point in Sec.~\ref{sec:algo-construction}, where we show that the BIT minimizes a certain MSE loss function. 
Meanwhile, we drop $\boldsymbol{z}$ in the notation when it is not needed.

\section{Decision tree fit as Fisher information optimization}\label{sec:algo-motivation}

Before we construct the new algorithm, it is instructive to interpret the weak learner of a traditional BDT algorithm in the context of Poisson counting experiments. 
It is defined by a maximum number $D$ of consecutive requirements on the vector of input features $\boldsymbol{x}$, collectively denoted by $\alpha_j$, that group the input data in no more than $2^D$ terminal nodes $j\in\mathcal{J}$. 
All terminal nodes are disjoint and the union of all $j\in\mathcal{J}$ equals the total feature space. 
The training minimizes a loss function with respect to $\alpha_j$ which is evaluated on the categorial input training data $\{y_i, \mathbf{x}_i\}_{i=1}^N$ . 
For binary classification, the labels can be chosen $y_i\in\{-1,1\}$ and the prediction is given by the majority 
$F_j=\textrm{sgn}\left(\sum_{i\in j}y_i\right)$ 
of the training labels in each terminal node $j$. 
The weak learner's binary prediction can then be written as
\begin{equation}
F(\mathbf{x})=\sum_{j\in \mathcal{J}}\mathds{1}_{\alpha_j}(\mathbf{x})F_j,
\end{equation}
 where $\mathds{1}_{\alpha_j}(\mathbf{x})=1$ if the feature vector $\mathbf{x}$ satisfies the requirements $\alpha_j$ of the terminal node $j$ and is zero otherwise. 
 In HEP, we call training data with $y_i=1$ ($y_i=-1$) signal (background) events and we denote their total number by s (b); the purity is defined as $\rho=s/(s+b)$ and the total yield is $\lambda=s+b$.
 During the training of the decision tree, 
 a loss function
 \begin{equation}
 \Delta_H = \lambda_1 H(\rho_1)+\lambda_2 H(\rho_2)-\lambda H(\rho)\label{eq:node-split}
 \end{equation}
is minimized to recursively split nodes.
Here, the primary node is characterized by ($\rho, \lambda$) and the secondary nodes by ($\rho_{1}, \lambda_{1}$) and ($\rho_{2}, \lambda_{2}$). 
A particular choice for $H$, the Gini impurity $H_G=\rho(1-\rho)$, leads to 
\begin{equation}
\Delta_{H_G} = -\left(\frac{1}{\lambda_1}+\frac{1}{\lambda_2}\right)^{-1}(\rho_1-\rho_2)^2\label{eq:node-split-gini}
\end{equation} and implies that optimal node splits result in maximally different secondary purities with a penalty for small yields. 

Equation~\ref{eq:node-split-gini} has an interesting interpretation in HEP, first explained in Ref.~\cite{Valassi:2019uhy}. 
The Fisher information matrix, defined as the variance of the score vector,
\begin{equation}
I_{ab}(\boldsymbol{\theta})=\mathds{E}_{\boldsymbol{\theta}}\left[\frac{\partial}{\partial \theta_a}\log p(\boldsymbol{x}|\boldsymbol{\theta})\frac{\partial}{\partial \theta_b}\log p(\boldsymbol{x}|\boldsymbol{\theta})\right],\label{eq:fisher}
\end{equation} 
has many useful applications in parameter estimation and regression problems.
The Cram\'er-Rao bound~\cite{RaoCR,CrammerH}, e.g., states that the variance of an unbiased scalar estimator $\hat\theta$ of $\theta$ satisfies
\begin{equation}
\textrm{var}(\hat\theta)\geq\frac{1}{I(\theta)},
\end{equation}
which we can use as an approximate estimator of its expected variance.

In binned HEP data analyses, we partition the kinematic phase space at the reconstruction level in disjoint regions and perform Poisson counting experiments on the resulting yields. 
For a single Poisson yield $\lambda(\theta)$, the Fisher information matrix becomes
\begin{equation}
I_{ab}^{(\lambda)}(\boldsymbol{\theta})=\frac{1}{\lambda}\frac{\partial\lambda}{\partial \theta_a}\frac{\partial\lambda}{\partial \theta_b}.
\end{equation}
Let us specify this expression to the simple case of a single-bin measurement of the signal cross section $\Delta\sigma_s$.
We can write $\lambda(\Delta\sigma_s)=b+s(\Delta\sigma_s)=b+\Delta\sigma_s\mathcal{L}=\frac{1}{\rho}\Delta\sigma_s\mathcal{L}$ and thus, 
\begin{equation}
I^{(\lambda)}_{\Delta\sigma_s} = \frac{1}{\lambda}\left(\frac{\partial\lambda}{\partial\Delta\sigma_s}\right)^2 = \frac{\rho\mathcal{L}}{\Delta\sigma_s}=\frac{1}{(\Delta\sigma_s)^2}\lambda \rho^2.\label{eq:FI-pois-meas}
\end{equation}
We can drop the irrelevant constant factor $(\Delta\sigma_s)^{-2}$ and compare Eq.~\ref{eq:FI-pois-meas} to the $\lambda H$ terms in Eq.~\ref{eq:node-split}. We find a criterion $H_F=-\rho^2$ will  maximize the Fisher information $I_{\Delta\sigma_s}$. 
Because $H_F$ and $H_G$ differ only by terms linear in $\rho$ and such contributions drop out in Eq.~\ref{eq:node-split}, we conclude that the training of a decision tree with $\Delta_{H_G}$ is equivalent to maximizing the Fisher information in a cross section measurement of the signal.

\section{Boosted Information Trees}\label{sec:algo-construction}

The interpretation stated at the end of Sec.~\ref{sec:algo-motivation} suggests to define a node split criterion for general $\lambda(\theta)$. 
We define a training loss in terms of the Fisher information of a single Poisson yield by generalizing to
\begin{equation}
    H_F(\lambda)=-\left.\left(\frac{1}{\lambda}\frac{\partial\lambda}{\partial \theta}\right)^2\right|_{\theta=\theta_0},
\end{equation}
leading to a loss function
\begin{eqnarray}
\Delta_{H_F} &=& \lambda_1 H_F(\lambda_1)+\lambda_2 H_F(\lambda_2)-\lambda H_F(\lambda)\nonumber\\
&=&-I^{(\lambda_1)}(\theta_0)-I^{(\lambda_2)}(\theta_0)+I^{(\lambda)}(\theta_0)\nonumber\\
&=& -\left.\left(\frac{1}{\lambda_1}+\frac{1}{\lambda_2}\right)^{-1}\left(\frac{\partial}{\partial \theta}\log \lambda_1 - \frac{\partial}{\partial \theta}\log \lambda_2\right)^2\right|_{\theta=\theta_0}\label{eq:node-split-fisher}.
\end{eqnarray}
This node split criterion thus maximizes the gain in the Fisher information of a scalar parameter $\theta$ of Poisson yields or, equivalently, the difference in the fractional rate of change $\partial_\theta\log\lambda(\theta)=\lambda'(\theta)/\lambda(\theta)$ of the secondary yields with same penalty term as in Eq.~\ref{eq:node-split-gini}. 
Note that a constant $\theta$-dependent factor, reflecting a $\theta$ dependence of the total cross section common to all training events, drops out from Eq.~\ref{eq:node-split-fisher}.
Therefore, either version of the augmented data in Sec.~\ref{sec:data} can be used.
The correspondence of Eq.~\ref{eq:node-split-gini} and Eq.~\ref{eq:node-split-fisher} can be understood by a loose analogy between $b\leftrightarrow\lambda(\theta)$ and $s+b\leftrightarrow\lambda(\theta+\Delta \theta)$ such that $\rho\leftrightarrow \lambda^{-1}\Delta\lambda/\Delta \theta$ for small $\Delta \theta$.
From Eq.~\ref{eq:node-split-fisher} we can thus determine the feature space partitioning $\mathcal{J}$ of a weak learner. 
It does, however, not determine the predicted value, which will become clear when we next consider boosting.

In classification, boosting minimizes a loss function $L(y, F(\mathbf{x}))$ by iteratively fitting weak learners to the pseudo-residuals of the preceding iteration step. The AdaBoost.M1 algorithm~\cite{adaBoost-original}, e.g., has been shown to minimize the exponential loss $\mathds{E}[\exp(-yF(x))]$ where $y$ denotes the true classification value~\cite{10.1214/aos/1016218223}.
The product $-yF(x)$ is negative for a correct classification and corresponds to $-\frac{\partial w}{\partial \theta}F(x)$ for an augmented training event $(w(\theta), \mathbf{x})$.
We therefore make the corresponding linear ansatz
\begin{equation}
F(\mathbf{x})=-\sum_{j\in \mathcal{J}}\mathds{1}_{\alpha_j}(\mathbf{x})F_{j}\left.\frac{\partial w}{\partial \theta}\right|_{\theta=\theta_0},
\end{equation}
where the constants $F_j$ are the weak learner's prediction for a new event that falls into the terminal node $j\in \mathcal{J}$. 
It is straightforward to verify that the choice 
\begin{equation}
    F_j=\frac{\partial}{\partial \theta} \log \sum_{i\in j} w_i(\theta)\bigg|_{\theta=\theta_0} = \frac{1}{\lambda_j}\left.\frac{\partial \lambda_j}{\partial \theta}
\right|_{\theta=\theta_0}\label{eq:Fj-choice}
\end{equation}
results in 
\begin{equation}
    \mathds{E}(F)=-\sum_{j\in \mathcal{J}}\frac{1}{\lambda_j}\left.\left(\frac{\partial \lambda _j}{\partial \theta}\right)^2\right|_{\theta=\theta_0}=-\sum_{j\in\mathcal{J}}I^{(\lambda_j)}(\theta_0)\label{eq:FI-loss}
\end{equation}
and we thus obtain a discriminator that attempts to partition the training data to maximize the cumulative Poisson Fisher Information of the partitioned training yields. 

Because terms $\mathcal{O}\left((\theta-\theta_0)^2\right)$ in the expansion of $w(\theta)$ drop out, we simplify the input data further to the weight and its derivative, 
\begin{equation}
\{\mathbf{x}_i,w_i,w'_i\}=\left\{\mathbf{x}_i, w_i(\theta_\textrm{0}),\left.\frac{\partial w_i}{\partial \theta}\right|_{\theta_{\textrm{0}}}\right\},\label{eq:training-data}
\end{equation}
evaluated at the parameter point where the discriminator is trained.
In Algorithm \ref{alg:weak-learner}, we present the pseudo-code for the fitting procedure of our decision-tree weak learner. 
It is an adaption of the well-known CART algorithm by Breiman et al.~\cite{breiman1984classification}, where the feature space in each recursion step is further divided by greedily selecting the dimension and cut value combination that maximizes a local gain. 
The main differences are that we maximize the Fisher information as node split criterion as described above and set the log-derivative of the node yield as prediction corresponding to a terminal node.
Overfitting is mitigated by enforcing a maximum tree depth~$D$ and a minimum number of events $N_{\textrm{min}}$ in a terminal node.
Since all events are partitioned among the nodes of each layer in the tree, the runtime of the algorithm scales with $\mathcal{O}(d D N \log N)$, where $d$ is the dimension of the features and $N$ the number of events. 
It can be efficiently implemented by sorting the events' features for each dimension separately and using cumulative sums over the weights and weight derivatives to find the locally optimal cut. 
The terminal nodes $j \in \mathcal J$ are encoded each as tuple consisting of the requirements $\alpha_j$ (selecting a subset of the data $\mathcal{D}_{\alpha} \subset \mathcal{D}$) and prediction $F_j$. 
They determine the final weak learner $F(\mathbf{x}) = \sum_{j \in \mathcal{J}} \mathds{1}_{\alpha_j}(\mathbf{x}) \cdot F_j$.
\begin{algorithm}[p]
    \setstretch{1.1}
	\KwData{Data set $\mathcal{D} = \{\mathbf{x}_i,w_i, w'_i\}_{i=1}^N$.}
	\KwIn{Tree depth $D$, minimum terminal node size $N_\mathrm{min}$}
	\KwOut{Weak learner $F$ \vspace{0.02cm}}
	
	$\alpha \gets (), Q \gets (\alpha), \mathcal{J} \gets \{\}, P \gets \{1, \dots, d\}$\;
	
	\While{$Q \ne \emptyset$}{
		$\alpha \gets \mathsf{pop}(Q)$\;
		
		$\bm \pi \gets \mathrm{arg\,sort}\ \mathbf{x} \in \mathcal{D}_{\alpha}$ \algorithmiccomment{sorted indices for each dimension $p \in P$}\;
		
		\uIf{$|\alpha| \le D \land |\mathcal{D}_{\alpha}| > 2N_{\textrm{min}}$}{
		    $K \gets \{N_{\textrm{min}},\dots, |\mathcal{D}_{\alpha}| - N_{\textrm{min}}\}$\; \algorithmiccomment{allowed cut points}
		    
			$p^*, k^* \gets \argmax_{p\in P, k\in K} \left[\frac{\left(\sum_{i=1}^k w'_{\bm \pi_{p, k}}\right)^2}{ \sum_{i=1}^k w_{\bm \pi_{p, k}}} + \frac{\left(\sum_{i=k+1}^{|\mathcal{D}_{\alpha}|} w_{\bm \pi_{p, k}}\right)^2}{\sum_{i=k+1}^{|\mathcal{D}_{\alpha}|} w_{\bm \pi_{p, k}}}\right]$\;
			
			$c \gets x_{\bm \pi_{p^*,k^*},p^*}$ from $\mathcal{D}_{\alpha}$\;
			
			\uIf{$c$ is a valid cut}{
				$\alpha_\mathrm{L} \gets \alpha \cup (p^*, \le, c)$\;
				
				$\alpha_\mathrm{R} \gets \alpha \cup (p^*, >, c)$\;
				
				$Q \gets Q \cup (\alpha_\mathrm{L}, \alpha_\mathrm{R})$\;
			}
			\Else{
				$\mathcal{J} \gets \mathcal{J} \cup \left(\alpha, \frac{\sum_{w' \in \mathcal{D}_{\alpha}} w'}{ \sum_{w \in \mathcal{D}_{\alpha}} w }\right)$\;
			}	
		}
		\Else{
			$\mathcal{J} \gets \mathcal{J} \cup \left(\alpha, \frac{\sum_{w' \in \mathcal{D}_{\alpha}} w' }{\sum_{w \in \mathcal{D}_{\alpha}} w }\right)$\;
		}		
	}
	\Return $F(\mathbf{x}) = \sum_{j \in \mathcal{J}} \mathds{1}_{\alpha_j}(\mathbf{x}) \cdot F_j$\;
	
	\caption{The $\mathsf{fit}$ procedure for the weak learner } 
	\label{alg:weak-learner}
\end{algorithm}
Choosing a learning rate $\eta$ and a number of boosting iterations $B$, the final algorithm is described in Algorithm~\ref{algo:bit}. 
In each iteration $b$, only a fraction $\eta f^{(b)}$ is retained from the weak learner for the cumulant prediction $F^{(b)}$. 
The replacement $w_i'\rightarrow w'_i -\eta w_iF^{(b-1)}$ implements the corresponding reduction in the local score estimates by reducing the weight derivatives for the next iteration.
\begin{algorithm}[t]
\KwData{Data set $\mathcal{D} = \{\mathbf{x}_i,w_i,w'_i\}_{i=1}^N$.}
\KwIn{Number of boosting iterations $B$, learning rate $\eta$}
 \KwOut{Boosted learner $F^{(B)}$ \vspace{0.02cm}}
$F^{(0)} \gets 0$
 \algorithmiccomment{initialize boosted learner}\;

\For{$b \gets 1,\hdots,B$}{

    $f^{(b)} \gets \mathsf{fit}\left(\left\{ \mathbf{x}_i, w_i, w'_i -\eta w_iF^{(b-1)}(\mathbf{x}_i) \right\}_{i=1}^{{N}}\right)$ \;
    
     $F^{(b)} \gets F^{(b-1)}+\eta f^{(b)}$\;
    
}
\caption{Boosted Information Tree} 
\label{algo:bit}
\end{algorithm}

Equation~\ref{eq:Fj-choice} can indeed be interpreted as an approximation of the score function in the current node of the decision tree, iteratively refined by the boosting algorithm. 
For any region $j$, we can compute the local mean of the score function of the PDF as 
\begin{equation}
    \frac{\mathds{E}_{\theta_0}\left(\mathds{1}_j(\boldsymbol{x})\cdot\partial_\theta\log p(\boldsymbol{x}|\theta)\Big|_{\theta=\theta_{\textrm{0}}}\right)}{\mathds{E}_{\theta_0}\left(\mathds{1}_j(\boldsymbol{x})\right)} = \partial_\theta\log\lambda\Big|_{\theta=\theta_{\textrm{0}}}-\partial_\theta\log\sigma\Big|_{\theta=\theta_{\textrm{0}}},
\end{equation}
which allows us to understand the discriminator output as
\begin{equation}
    F^{(b)}(\boldsymbol{x})\simeq t(\boldsymbol{x}|\theta_0)+\partial_\theta\log\sigma(\theta)\Big|_{\theta=\theta_{\textrm{0}}}.\label{eq:algo-result}
\end{equation}
The last term is a calculable bias term accounting for the $\theta$ dependence of the inclusive cross section. 
It does not change the performance of the discriminator in the analysis setting; in fact, the output can be normalized to the unit interval and this option is available in the implementation. 
If the augmented data set is at the PDF level (i.e. $\omega_i$ instead of $w_i$ without the factor $\sigma(\theta)$ as discussed in Sec.~\ref{sec:data}), the last term in Eq.~\ref{eq:algo-result} is absent.

Training simulation is needed only for a single reference parameter point which can, but need not be at $\theta_0$. 
The discriminator learns only from the leading term in the expansion around the reference point in parameter space;
contributions $\mathcal{O}\left((\theta-\theta_0)^2\right)$ do not enter Algorithm~\ref{alg:weak-learner}.
In this linear approximation, the likelihood can be written as 
\begin{equation}
    p(\boldsymbol{x}|\theta)=\frac{1}{Z(\theta)}p(t(\boldsymbol{x}|\theta_0)|\theta_0) \exp(t(\boldsymbol{x}|\theta_0)\cdot(\theta-\theta_0)),
\end{equation}
which is an exponential family. 
Therefore, the score $t(\boldsymbol{x}|\theta)$ is a sufficient statistic for $p(x|\theta)$, learned by the algorithm~\cite{Brehmer:2018eca}.

It is interesting to derive the node split criterion from a different starting point. 
Let us disregard the Fisher information and instead regress in the score. 
For a given node split at a value $\alpha$, we denote the regressed values in the nodes, defined by $x<\alpha$ and $x\geq\alpha$, as $F_L$ and $F_R$, respectively. 
With an ansatz for the weak regressor $\hat t(x|\alpha,F_L,F_R)=F_L\Theta(x<\alpha)+F_R\Theta(x\geq\alpha)$, we can minimize the MSE loss 
\begin{equation}
    \textrm{MSE}(\alpha, F_L, F_R)=\mathds{E}_{\theta_0}\left[\left(\frac{1}{\omega(\theta)}\frac{\partial\omega}{\partial\theta}\Big|_{\theta=\theta_{\textrm{0}}}-\hat t(x|\alpha,F_L,F_R)\right)^2\right].\label{eq:loss-regression}
\end{equation}
The equations $\frac{\partial\textrm{MSE}}{\partial F_L}=\frac{\partial\textrm{MSE}}{\partial F_R}=0$ can be solved for $F_L$ and $F_R$ and, dropping terms independent of the threshold $\alpha$, we arrive at
\begin{eqnarray}
    \textrm{MSE}(\alpha)&=&-\frac{1}{\lambda_L(\alpha)}\left(\frac{\partial\lambda_L(\alpha)}{\partial\theta}\right)^2\Bigg|_{\theta=\theta_{\textrm{0}}}-\frac{1}{\lambda_R(\alpha)}\left(\frac{\partial\lambda_R(\alpha)}{\partial\theta}\right)^2\Bigg|_{\theta=\theta_{\textrm{0}}}\nonumber\\
    &=&-I^{(\lambda_L(\alpha))}(\theta_0)-I^{(\lambda_R(\alpha))}(\theta_0),\label{eq:fisher-loss-fromchi2}
\end{eqnarray}
where $\lambda_{L(,R)}=\sum_{i\in L(,R)}\omega_i(\theta)$.
The loss in Eq.~\ref{eq:loss-regression} is of the general form of Eq.~27 in Ref.~\cite{Brehmer:2018eca} which implies we regress on the true score $t(\boldsymbol{x}|\theta)$ despite the augmented data being conditional on $\boldsymbol{z}$. 
Equation~\ref{eq:fisher-loss-fromchi2} shows that regressing in the MSE loss is equivalent to maximizing the cumulative Fisher information as a function of the threshold value $\alpha$, as is done in Eq.~\ref{eq:FI-loss} for the BIT. 
While the algorithm does approximate the score, there is no numerical minimization needed for the regressed value. 
Computationally, the algorithm behaves as a BDT. 

We note that Eq.~\ref{eq:loss-regression} suggests to extend the algorithm to estimate higher derivative terms of the likelihood, starting with the quadratic term, neglected here. Finally, in the case of more than one parameter, we can also expand the true likelihood ratio, i.e., the optimal test-statistic according to the Neyman-Pearson lemma, as 
\begin{equation}
r(\boldsymbol{x}|\boldsymbol{\theta},\boldsymbol{\theta}_{\textrm{ref}})=1+(\boldsymbol{\theta}-\boldsymbol{\theta}_{\textrm{ref}})\cdot\boldsymbol{t}(\boldsymbol{x}|\boldsymbol{\theta}_{\textrm{ref}}) +\mathcal{O}(\boldsymbol{\theta}-\boldsymbol{\theta}_{\textrm{ref}})^2.
\end{equation}
It implies that the combination  $(\theta_a-\theta_{a,\textrm{ref}}) t_a(\boldsymbol{x}|\boldsymbol{\theta}_{\textrm{ref}})$ is optimal in the linear approximation for a simple hypothesis test of $\boldsymbol{\theta}$ against $\boldsymbol{\theta}_{\textrm{ref}}$. The components $t_a(\boldsymbol{x}|\boldsymbol{\theta})$ of the score vector can be separately estimated by BITs, trained with the corresponding partial derivatives of the weight functions, i.e., $\partial_a w_i(\boldsymbol{\theta})|_{\boldsymbol{\theta}_{\textrm{ref}}}$ instead of $w'$ in Eq.~\ref{eq:training-data}.
We leave these developments to future work.

\section{Fitting analytic toy models}\label{sec:toys}

\begin{figure}[p]
    \centering

    \includegraphics[width=0.32\textwidth]{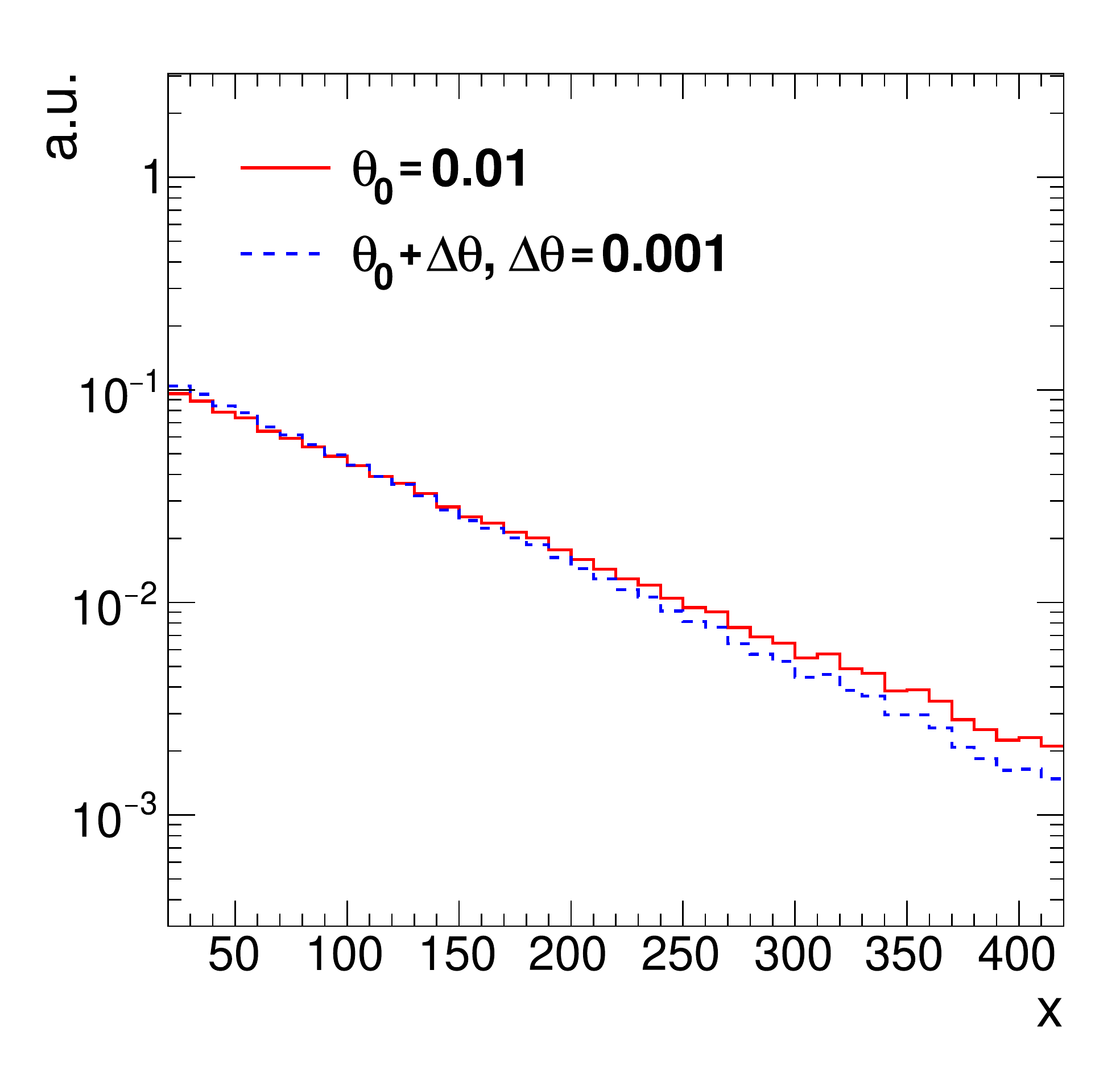}
    \includegraphics[width=0.32\textwidth]{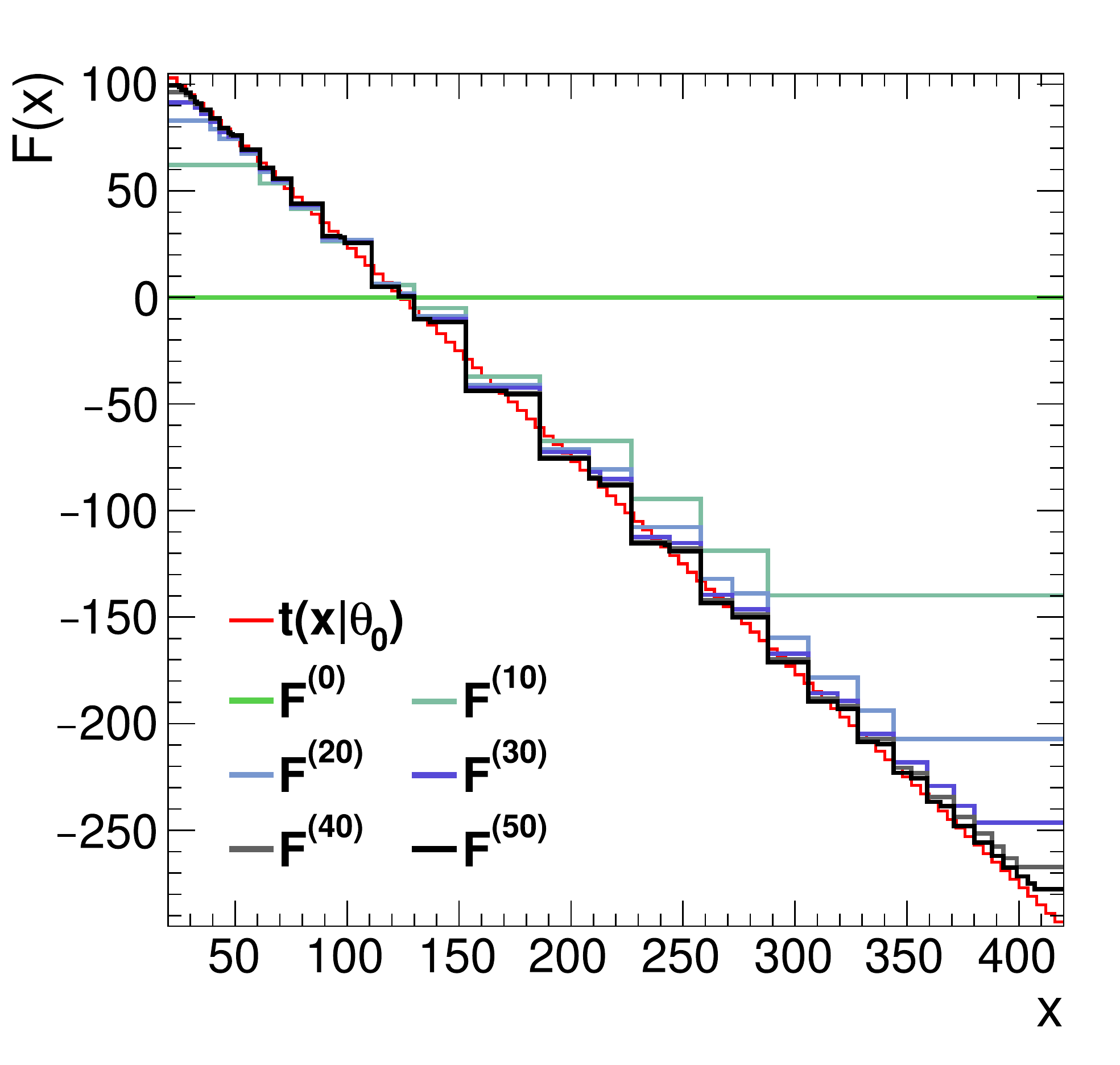}
    \includegraphics[width=0.32\textwidth]{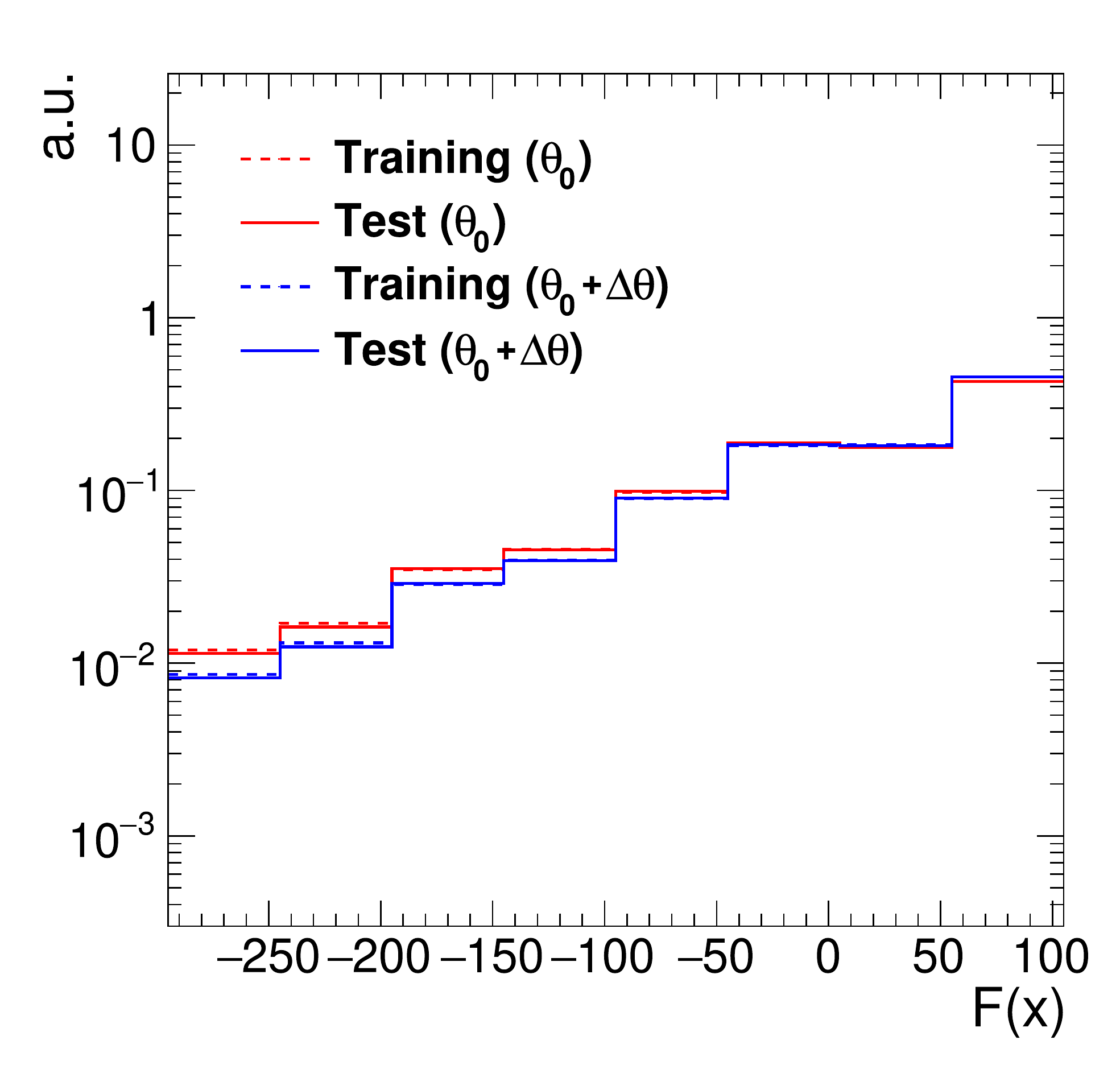}\\

    \includegraphics[width=0.32\textwidth]{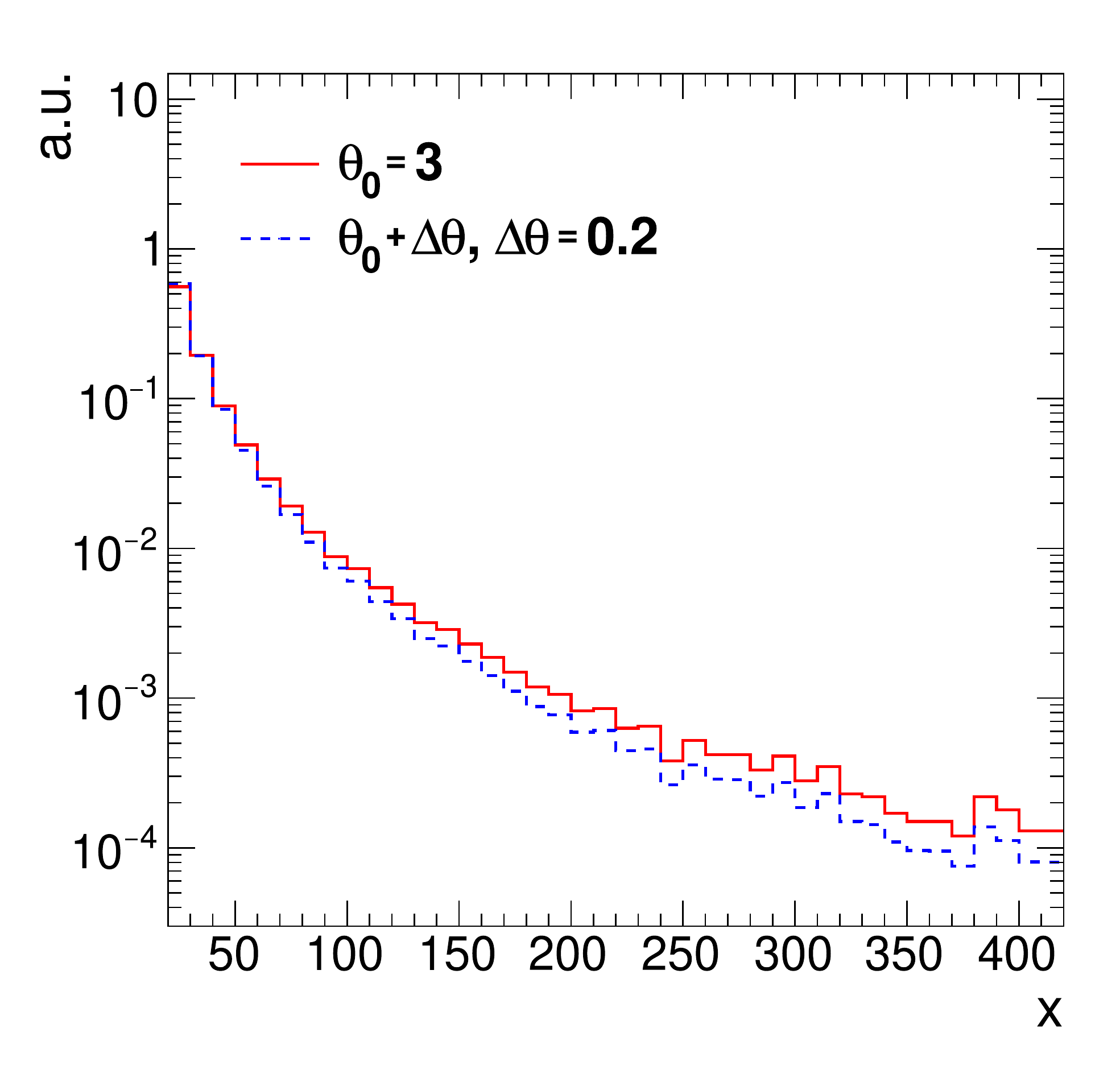}
    \includegraphics[width=0.32\textwidth]{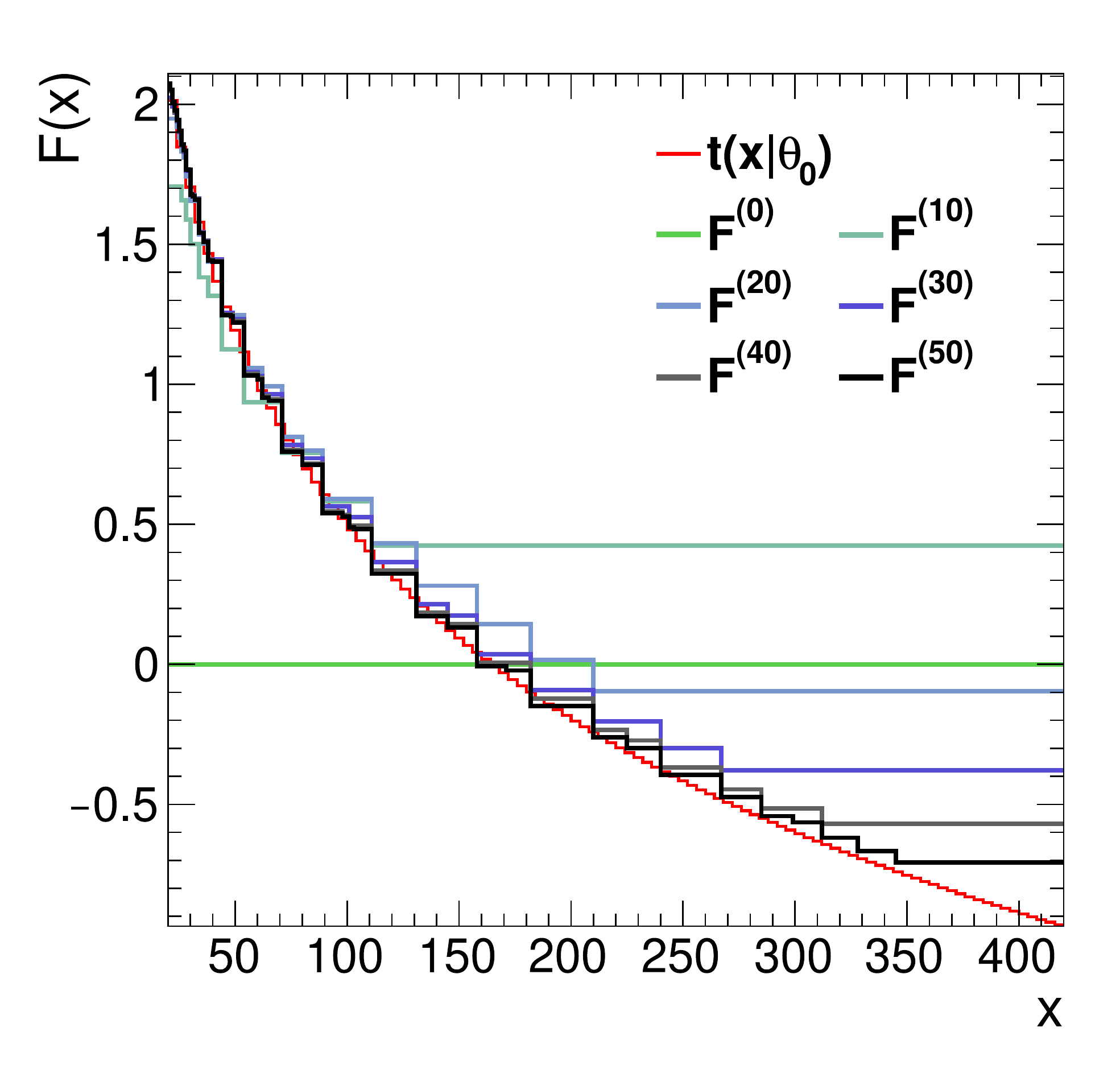}
    \includegraphics[width=0.32\textwidth]{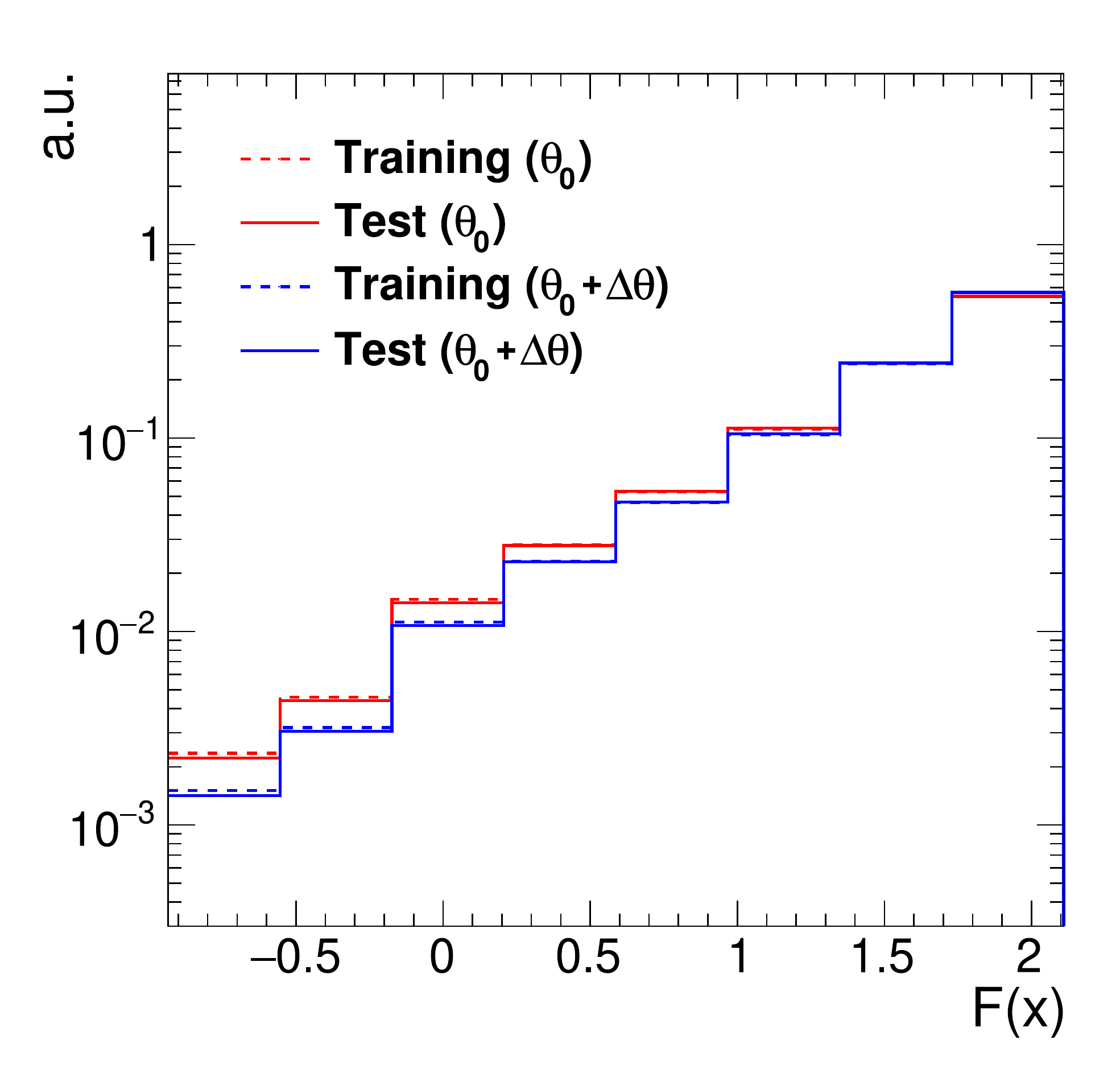}\\

    \includegraphics[width=0.32\textwidth]{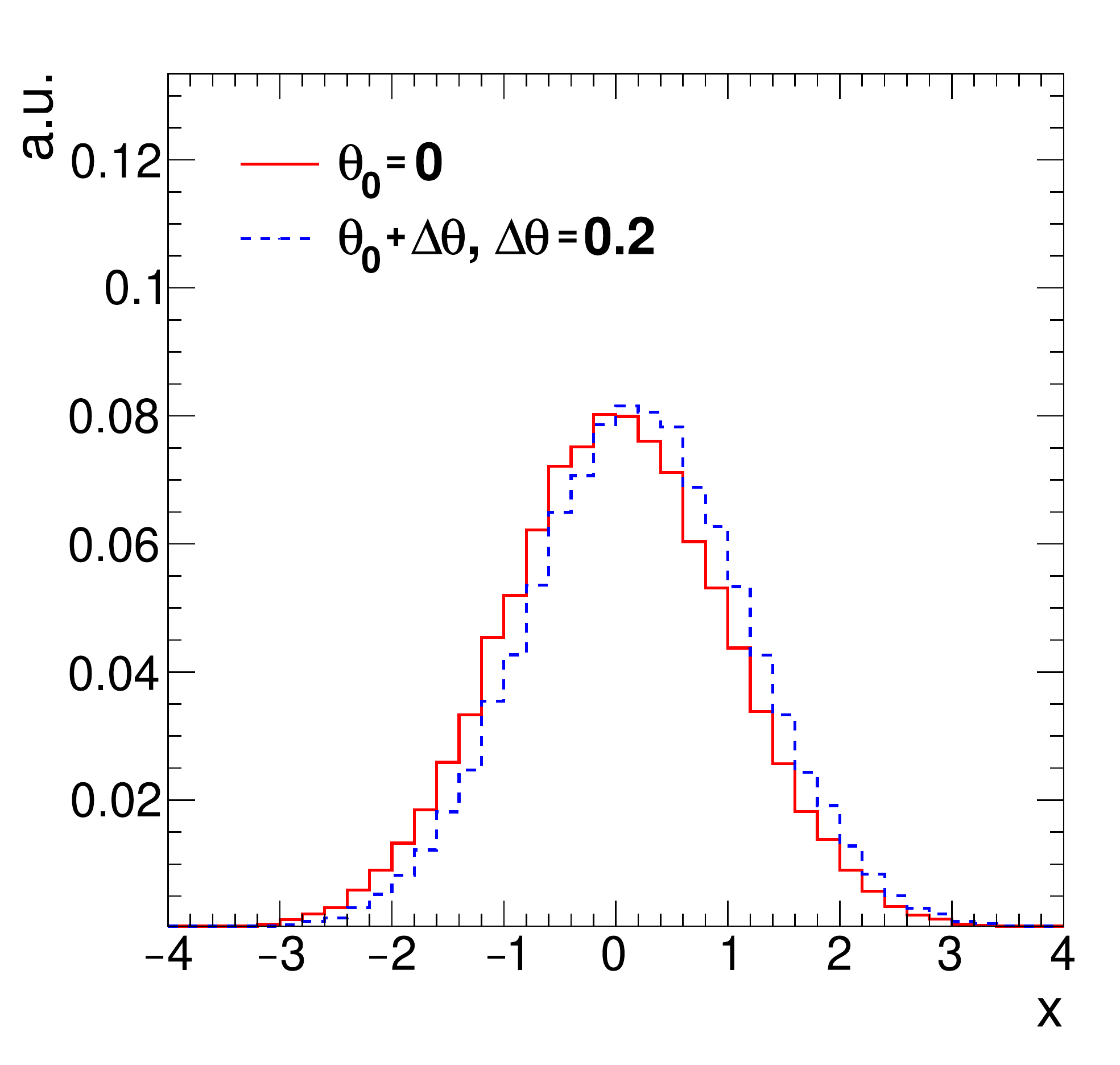}
    \includegraphics[width=0.32\textwidth]{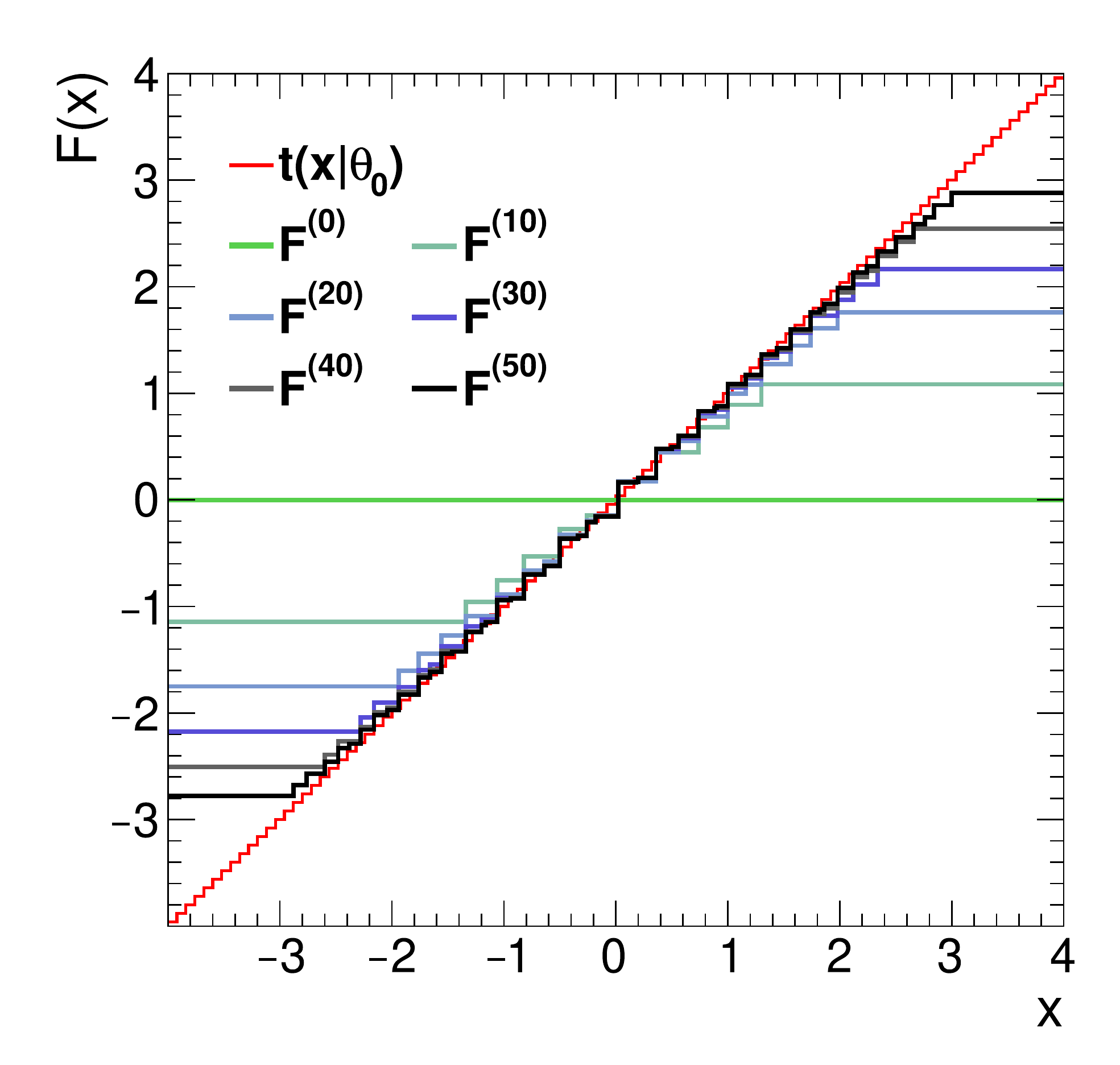}
    \includegraphics[width=0.32\textwidth]{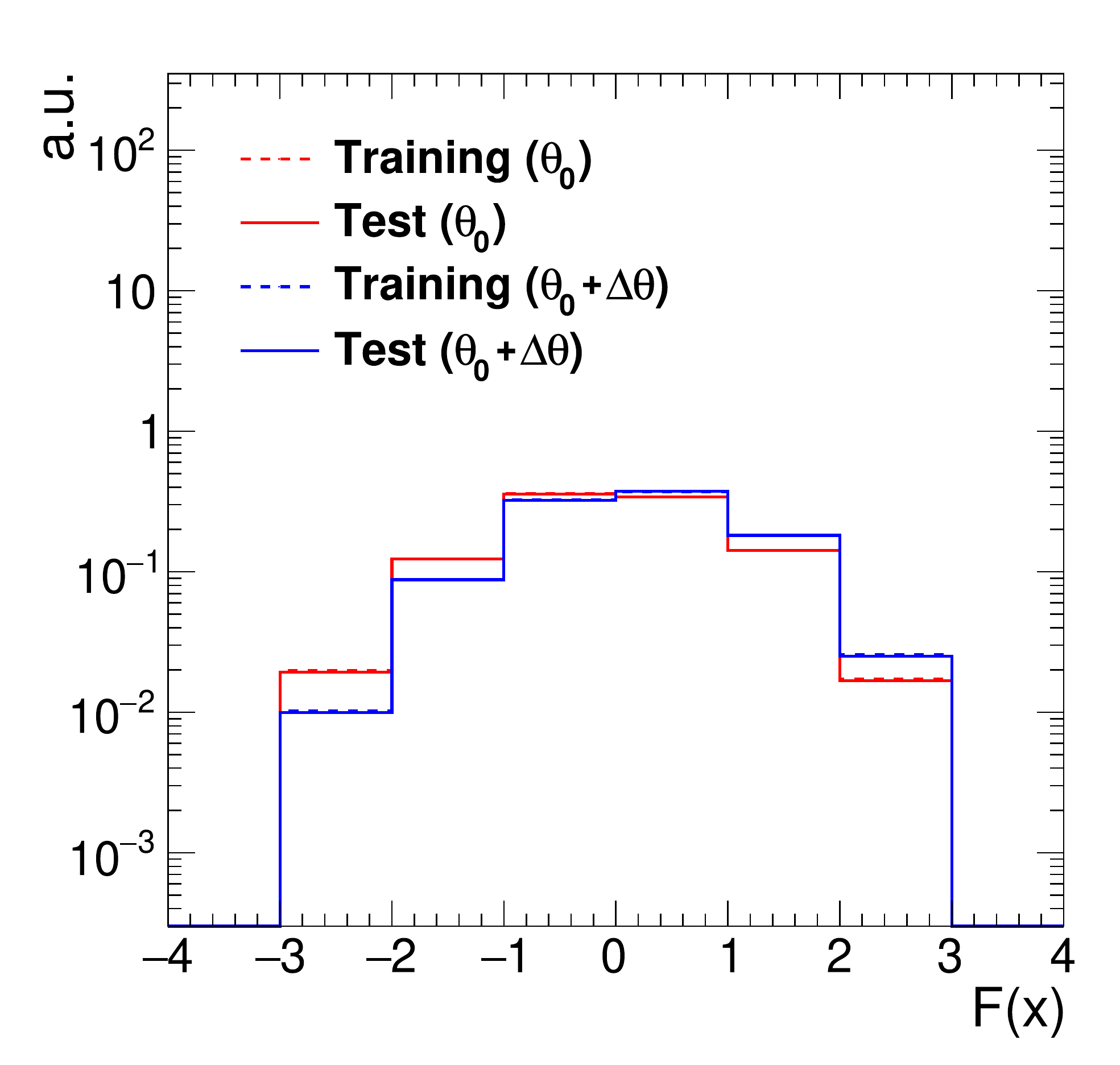}\\

    \includegraphics[width=0.32\textwidth]{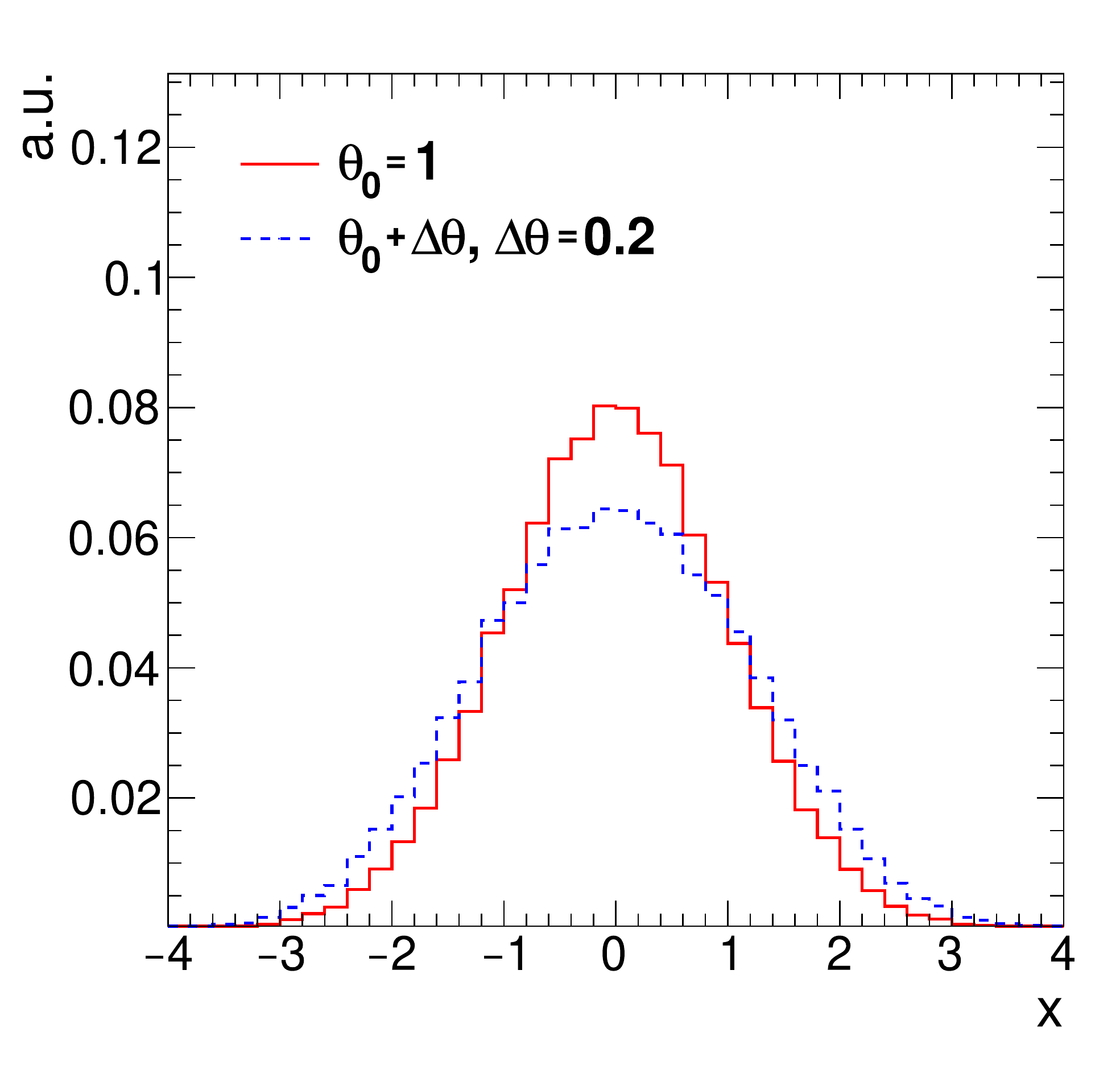}
    \includegraphics[width=0.32\textwidth]{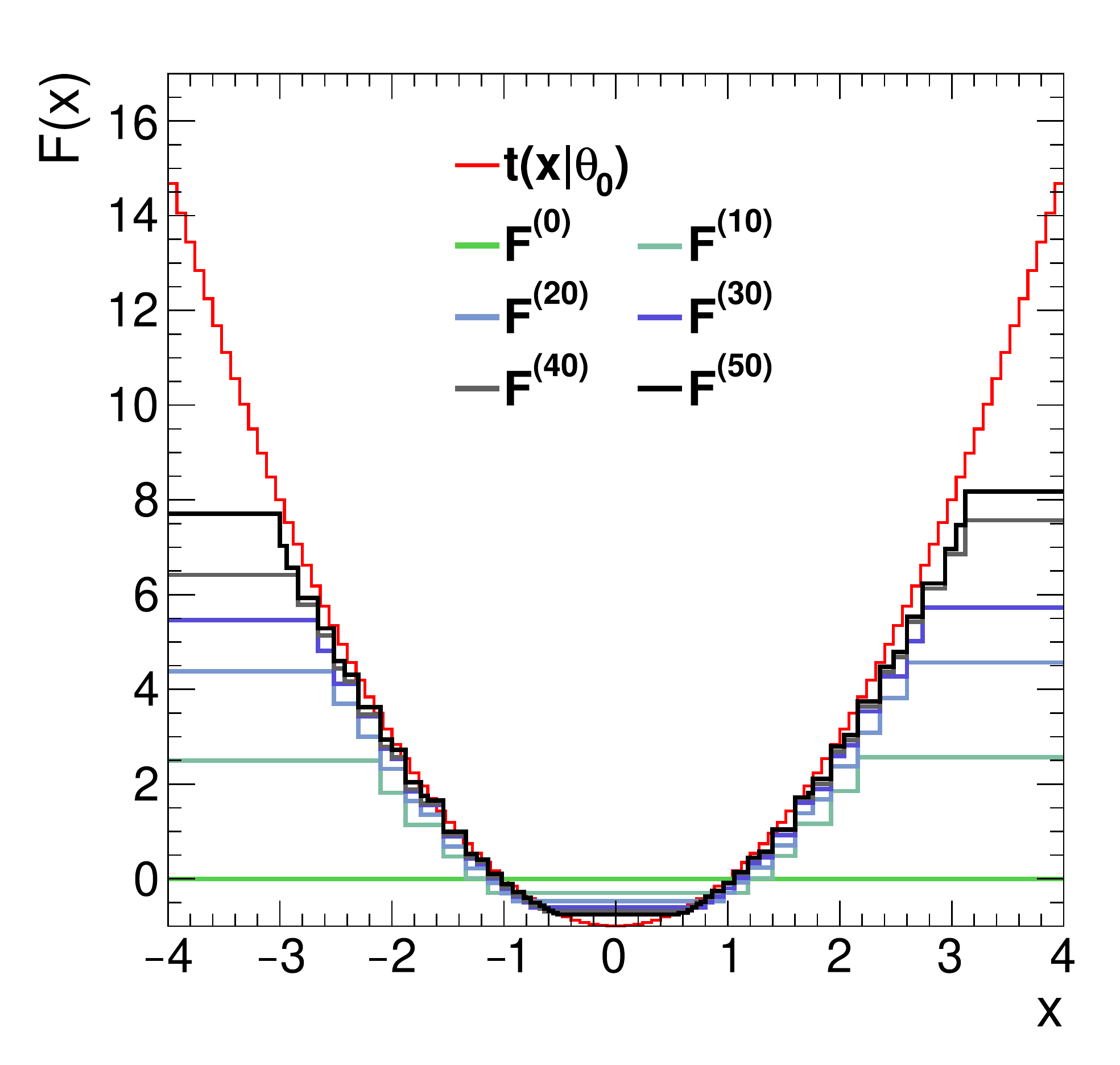}
    \includegraphics[width=0.32\textwidth]{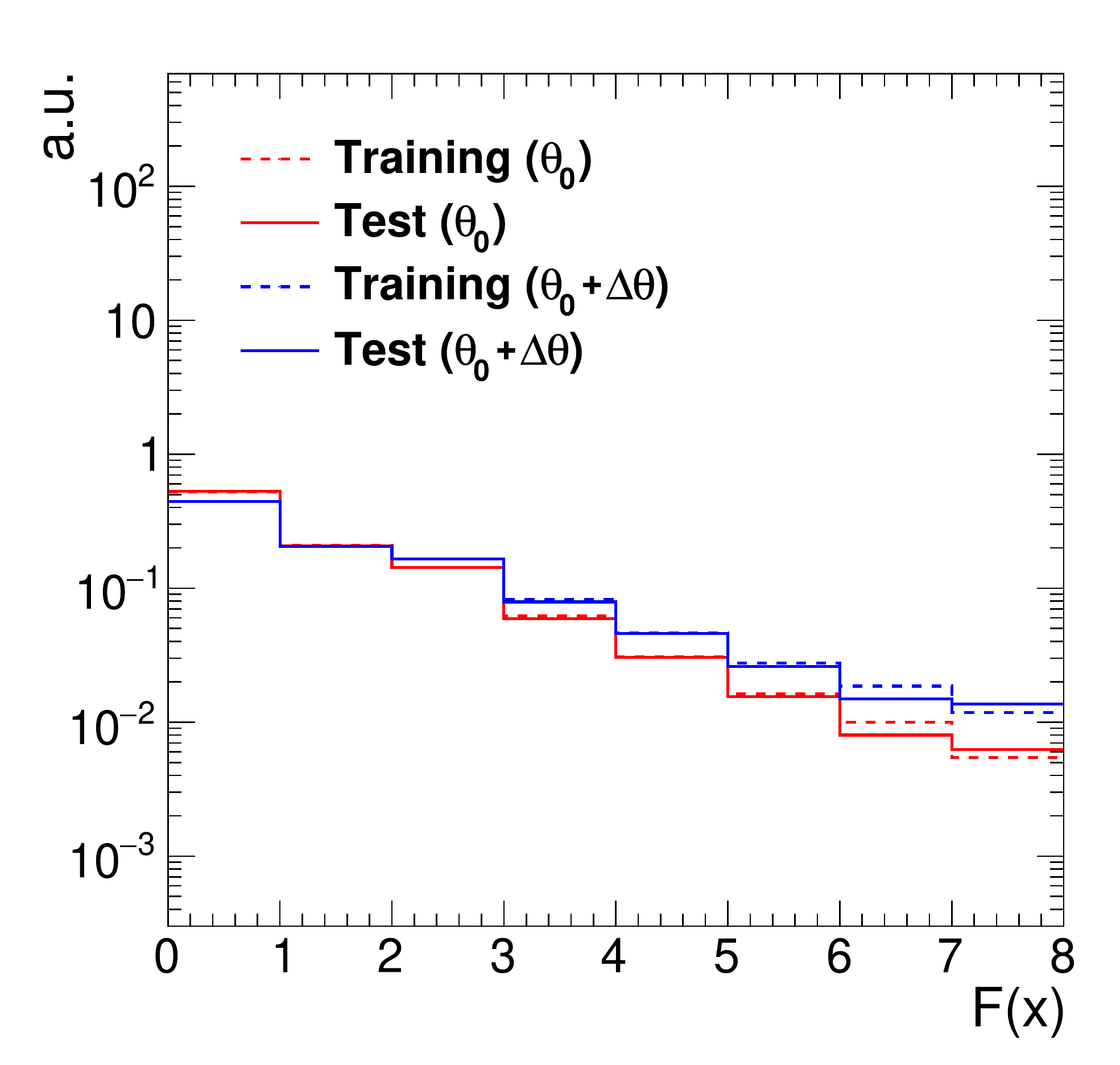}

    \caption{The PDF for toy models in Table~\ref{tab:toys} for $\theta_0$ and, for comparison, for an alternative parameter $\theta_0+\Delta\theta$~(left). Analytically computed score function $t(x,\theta_0)$ and the output of the BIT algorithm for different numbers of boosting iterations $b$~(center). Distributions of the predicted score function for the training data set and a statistically independent test data set for $\theta_0$ and $\theta_0+\Delta\theta$~(right). From top to bottom, the exponential, the power-law, the Gaussian-mean, and the Gaussian-width model are shown.  }\label{fig:toys}
\end{figure}

We now test the BIT on a variety of analytic toy models listed in Table~\ref{tab:toys}. 
We choose an exponentially falling PDF, a power-law model, a Gaussian model where the parameter is the mean, a Gaussian model where the parameter is the width, and a mixture model. 
In Fig.~\ref{fig:toys}~(left), we show the PDFs that we use to generate a data set consisting of $10^5$ events in each case. 
In Fig.~\ref{fig:toys}~(center), we compare the theoretical calculation with the predicted score at different numbers of boosting iterations $b$. 
The prediction nicely approximates the theoretically calculated score functions $t(\boldsymbol{x}|\theta)$ for increasing $b$. 
The distributions of the predicted score for the training data sets and statistically independent test data sets of the same size, evaluated for $\theta_0$ and a nearby parameter point $\theta_0+\Delta\theta$, are shown in Fig.~\ref{fig:toys}~(right). No significant overtraining is observed in the predicted distributions.

\begin{figure}[p]
      \centering
      \includegraphics[width=0.32\textwidth]{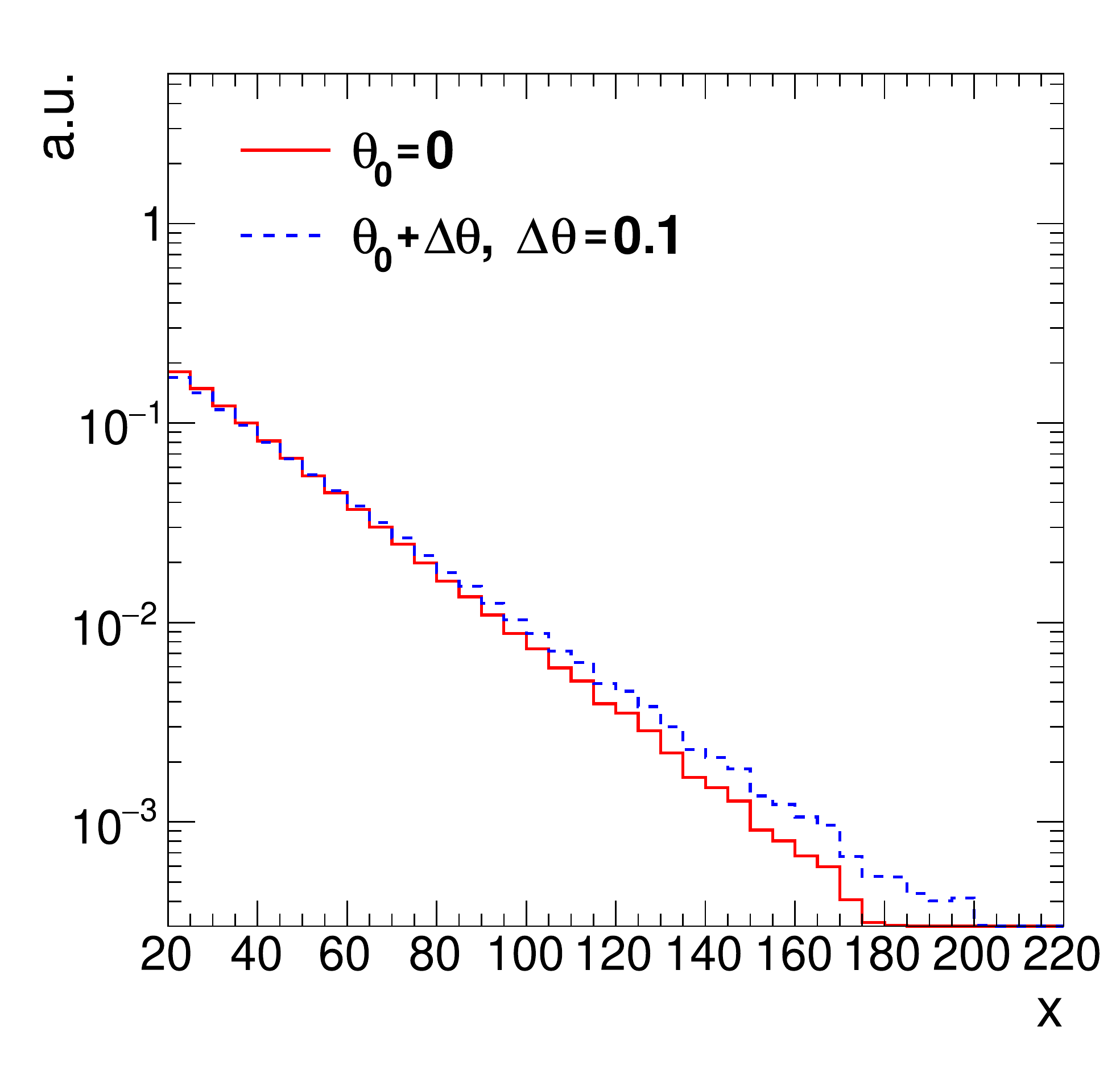}
      \includegraphics[width=0.32\textwidth]{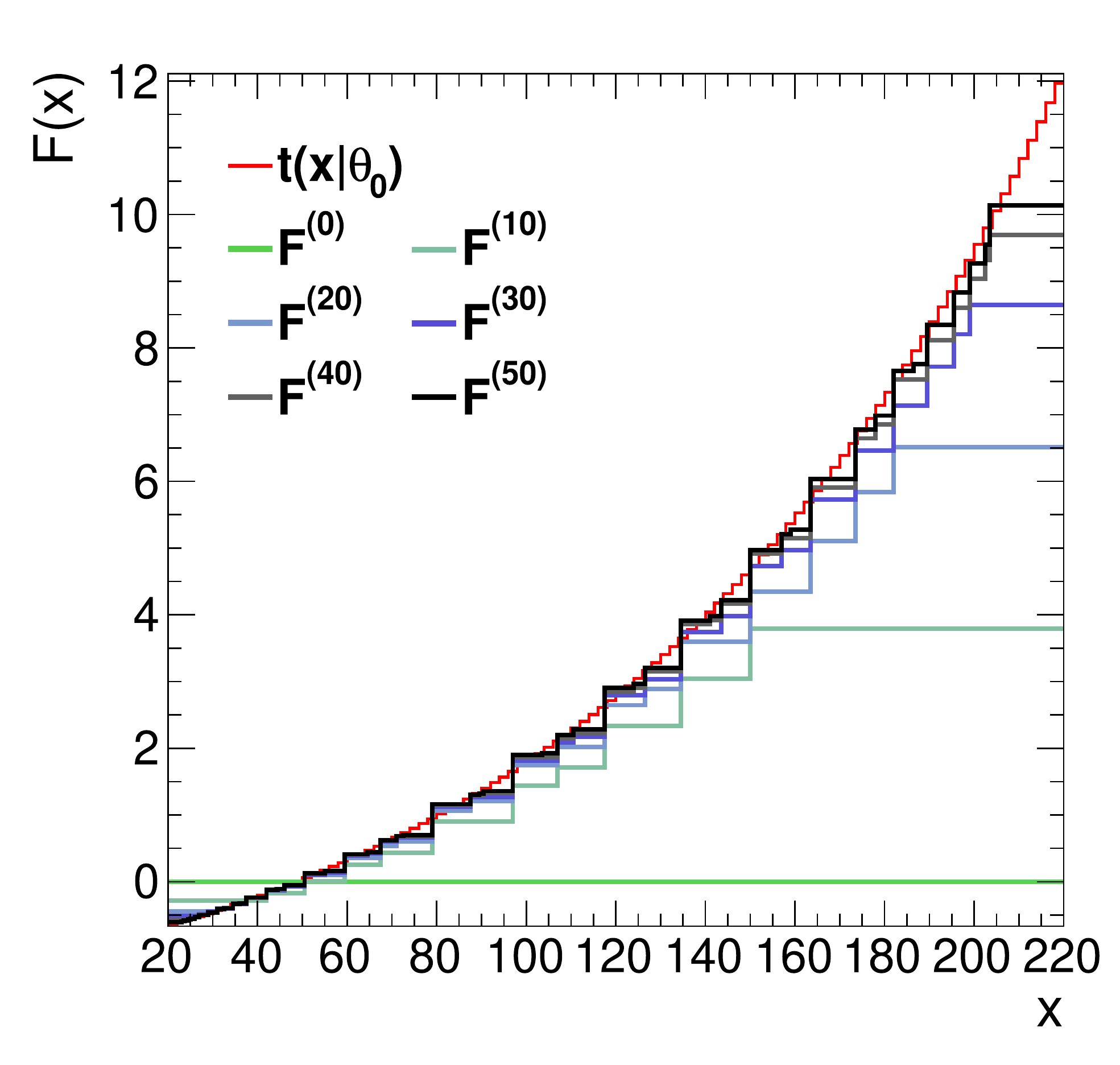}
      \includegraphics[width=0.32\textwidth]{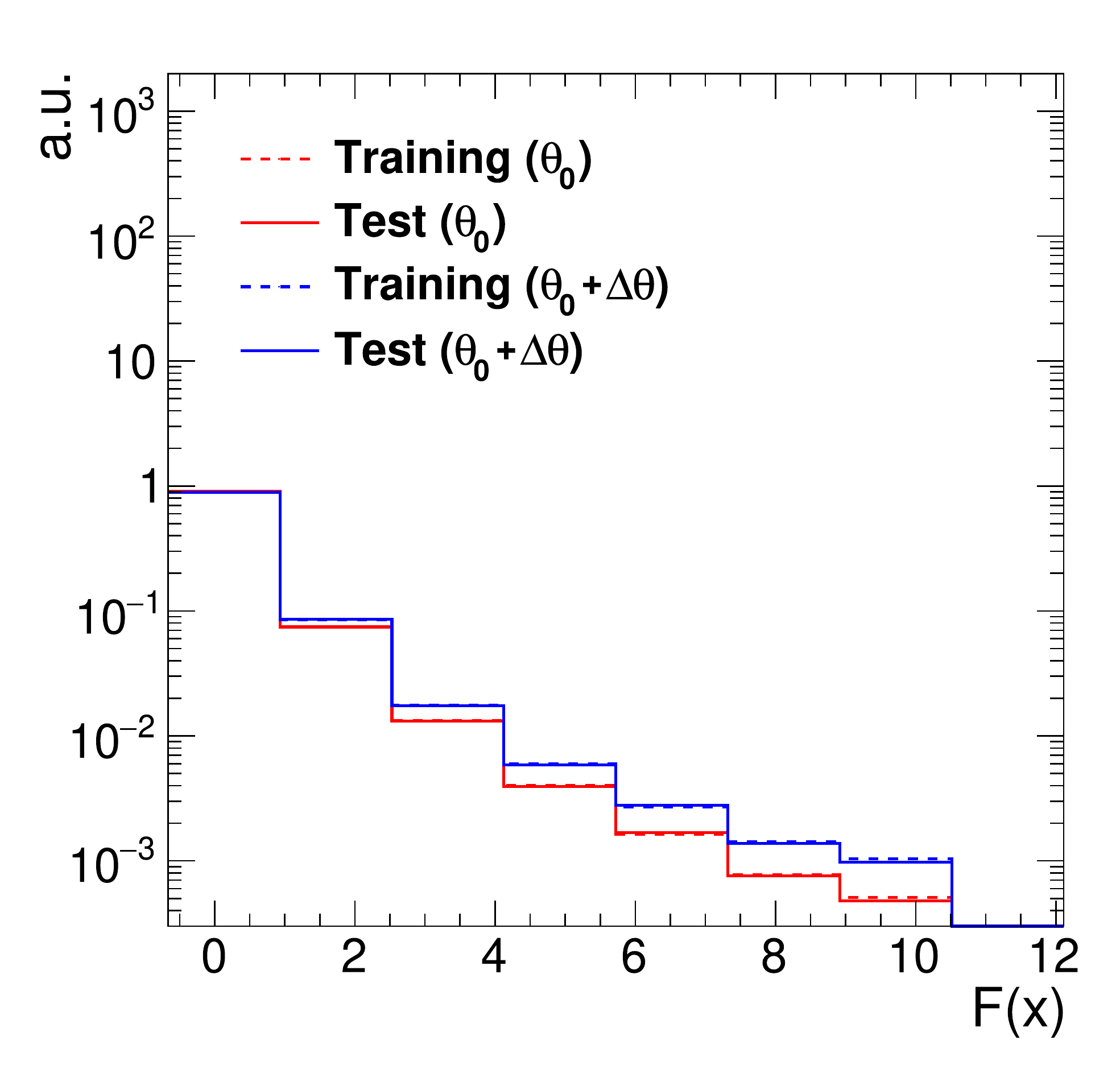}
      \caption{The PDF for the mixture model in Table~\ref{tab:toys} for $\theta_0$ and, for comparison, for an alternative parameter $\theta_0+\Delta\theta$~(left). Analytically computed score function $t(x,\theta_0)$ and the output of the BIT algorithm for different numbers of boosting iterations $b$~(center). Distributions of the predicted score function for the training data set and a statistically independent test data set for $\theta_0$ and $\theta_0+\Delta\theta$~(right).   }\label{fig:mixture}
  \end{figure}

In Fig.~\ref{fig:mixture}, we show the same distributions for a mixture of two exponential PDFs with different parameters $\alpha_{1,2}$,
\begin{equation}
    p(x|\theta)\propto\left(\exp(-\alpha_1(x-x_0))+\theta \exp(-\alpha_2(x-x_0))\right)^2.
\end{equation}
The dependence on the mixing parameter $\theta$ is broadly reminiscent of the structure in Eq.~\ref{eq:poly-xsec} while the choice of the exponential is arbitrary. The same conclusions as for the other toy models hold, raising the prospect of good performance in realistic EFT scenarios. 

{\renewcommand{\arraystretch}{1.2}
\begin{table}
\caption{PDF $p(x|\theta)$, score function $t(x|\theta)$, support of the PDF, constant parameters, and $\theta_0$ for the one-dimensional toy models.}
\label{tab:toys}
\begin{center}
\resizebox{11.5cm}{!}{
 \begin{tabular}{p{0.9\textwidth} c } 
  Exponential \\\hline
 \multirow{2}{*}{\begin{minipage}{3cm}\begin{eqnarray}p(x|\theta) &=& \theta e^{-\theta (x-x_0)},\;\;x>x_0, \;\;x_0=25,\;\; \theta_0=0.01 \nonumber\\t(x|\theta) &=& \frac{1}{\theta}-(x-x_0)\nonumber\end{eqnarray}\end{minipage}}\\\\\\\\
   Power law \\\hline
 \multirow{2}{*}{\begin{minipage}{3cm}\begin{eqnarray}p(x|\theta) &=& \frac{\theta-1}{x_0} \left(\frac{x}{x_0}\right)^{-\theta},\;\;x>x_0, \;\;x_0=100, \;\;\theta_0=3 \nonumber\\t(x|\theta) &=& \frac{1}{\theta-1}-\log\frac{x}{x_0}\nonumber\end{eqnarray}\end{minipage}}\\\\\\\\\\
    Gaussian mean\\\hline
 \multirow{2}{*}{\begin{minipage}{3cm}\begin{eqnarray}p(x|\theta) &=& \frac{1}{\sqrt{2\pi}\sigma}\exp\left(-\frac{1}{2}\left(\frac{x-\theta}{\sigma}\right)^2\right), \;\;\sigma=1, \;\;\theta_0=0 \nonumber\\t(x|\theta) &=& \frac{x-\theta}{\sigma^2}\nonumber\end{eqnarray}\end{minipage}}\\\\\\\\\\
    Gaussian width\\\hline
 \multirow{2}{*}{\begin{minipage}{3cm}\begin{eqnarray}p(x|\theta) &=& \frac{1}{\sqrt{2\pi}\theta}\exp\left(-\frac{1}{2}\left(\frac{x-\mu}{\theta}\right)^2\right), \;\;\mu=0, \;\;\theta_0=1 \nonumber\\t(x|\theta) &=& \frac{1}{\theta}\left(\left(\frac{x-\mu}{\theta}\right)^2-1\right)\nonumber\end{eqnarray}\end{minipage}}
  \\\\\\\\\\
     Mixture\\\hline
  \multirow{2}{*}{\begin{minipage}{3cm}\begin{eqnarray}p(x|\theta) &=& \left(\frac{1}{2\alpha_1}+\frac{2\theta}{\alpha_1+\alpha_2}+\frac{\theta^2}{2\alpha_2}\right)^{-1} \left(e^{-\alpha_1 (x-x_0)}+\theta e^{-\alpha_2 (x-x_0)}\right)^2 \nonumber\\
  t(x|\theta) &=& \frac{2}{\theta+e^{-(x-x_0)(\alpha_1-\alpha_2)}}-2\frac{\theta \alpha_1(\alpha_1+2\alpha_2)+2\alpha_1\alpha_2}{\alpha_2(\alpha_1+\alpha_2)+4\alpha_1\alpha_2\theta+\alpha_1(\alpha_1+\alpha_2)\theta^2}\nonumber\\ \alpha_1&=&2\alpha_2=0.02, \;\;x_0=20, \;\;\theta_0=0\nonumber\end{eqnarray}\end{minipage}}
\end{tabular}
}
\end{center}
\end{table}
}

\begin{figure}
    \centering
        \includegraphics[width=0.48\textwidth]{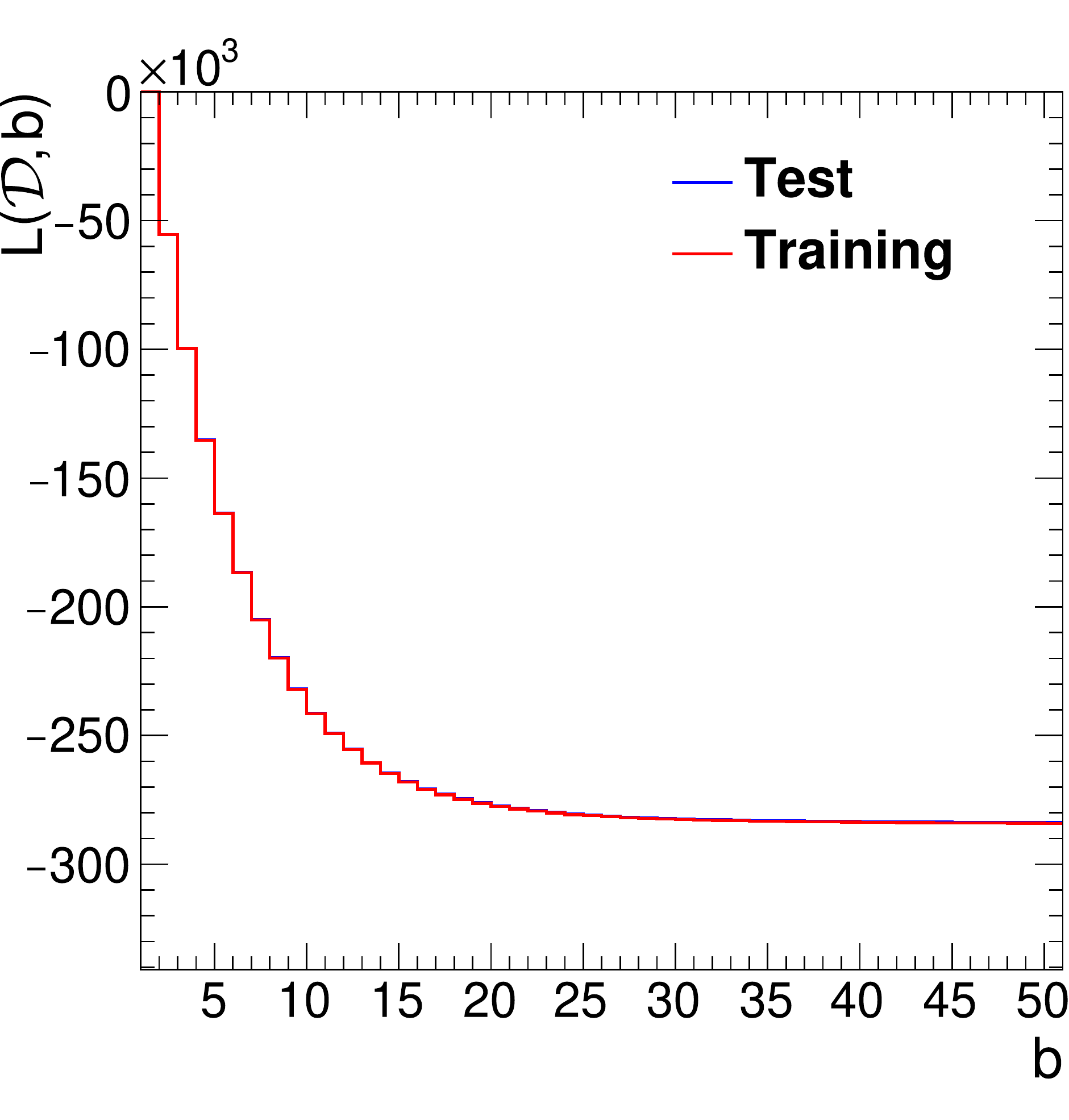}
    \hfill
    \includegraphics[width=0.48\textwidth]{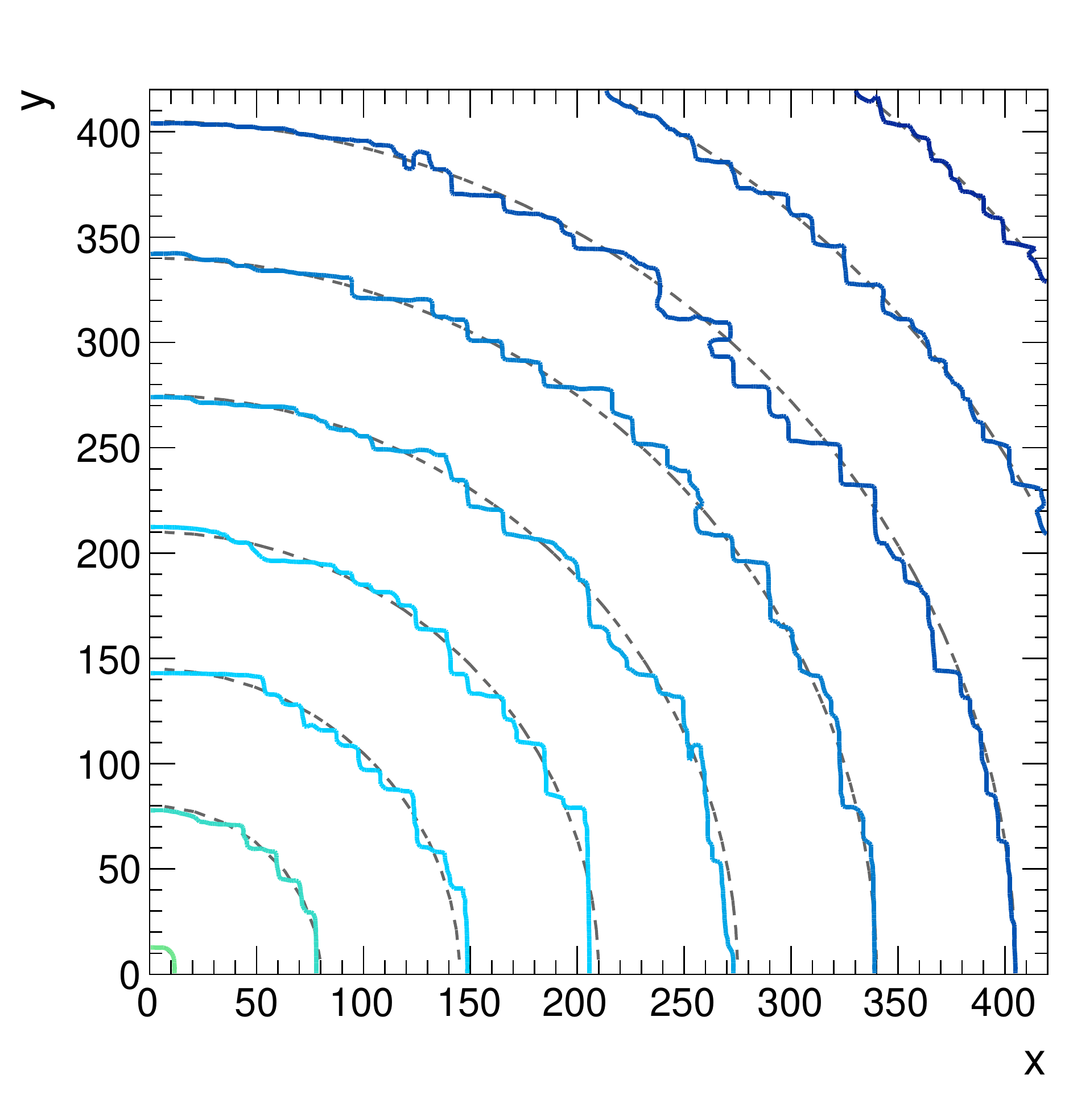}
    \caption{Overtraining measure $L(\mathcal{D},b)$ as a function of the boosting iteration~(left). There is no significant difference between the training and the statistically independent test data set. 
    Contour lines of the analytically computed score of the two-dimensional exponential model~(dashed lines) and the colored contour lines of the predicted score~(right). }\label{fig:multi-D}
\end{figure}

Overtraining can be monitored by evaluating the loss defined as     
 \begin{equation}
    L(\mathcal{D}, b)= -\sum_{i\in\mathcal{D}}\left(\frac{\partial\omega_i(\theta)}{\partial\theta}F^{(b)}(\mathbf{x}_i)\right)
 \end{equation}
 for the training and a statistically independent test data set. In Fig.~\ref{fig:multi-D}~(left) we show its evolution as a function of the boosting iteration. After about 30 iterations, the loss no longer reduces. A statistically independent data set shows the exact same behavior. The reason for the negligible overtraining lies in the homogeneity of the training data set in the toy case which is sampled from a simple analytic PDF. The algorithm cannot place thresholds based on statistically insignificant fluctuations in the training data set as in BDT classification: There are no signal and background events with independent stochastics. We expect the overtraining situation to deteriorate slightly when the variance of the $w_i$ is large in the training data set. 
In an extreme situation where $w_i'/w_i$ takes only two values, we expect the overtraining situation to resemble the classification case. In all studies conducted so far, a simple requirement $N\geq N_\textrm{min}$ on the events in each node of any of the weak learners is found to be sufficient to suppress significant overtraining. 
Further measures to mitigate overtraining should be informed by studies with realistic SM-EFT models that we leave to future work.

Because of the similarity of the BIT algorithm to BDT classification, we expect that other well-known differences of tree-boosting and neural-network based strategies prevail for the BIT.
Besides the reduced risk of overtraining already discussed, these include, for example, a generally faster execution time of the trained estimator, no intrinsic limitation in the approximation of large gradients in feature space, and a protection against extrapolation to regions in feature space without training data. 

Finally, we briefly test the performance in models with higher feature dimensions. Adding up to 25 random variables per feature does not affect the output of the one-dimensional analytic models in any way. 
A more challenging test of the rectangularly selecting boosted learners is to disperse the discriminating information in a non-linear way.
We consider a two-dimensional exponential model and train a BIT on $10^6$ events with $B=100$.
A comparison of the theoretically calculated contour lines of the score function and the contour lines of the regressed score is shown in Fig.~\ref{fig:multi-D}~(right). The main features are very well described by the discriminator. 
The test has been repeated with up to 5-dimensional variants of exponential and power-law models.
After moderate adjustments of training statistic and $B$, the fit quality to the score function is similar as in the one-dimensional cases.

We leave possible refinements of the algorithm to further mitigate the overtraining, a possible implementation in a versatile generic tree-boosting framework such as XGBoost~\cite{xgboost}, as well as a potential extension to quadratic terms in the likelihood expansion to future work.

\section{Conclusions and outlook}\label{sec:conclusion}

We presented the ``Boosted Information Tree'' algorithm, a boosted tree-based discriminator that optimizes the Fisher information in the node split of the weak learner. 
It regresses on the true score of the model at the computational cost of a decision tree.
It provides a sufficient statistic in the linear approximation around the reference point in parameter space where the discriminator is trained and can be used to improve analysis strategies for measuring EFT parameters.
Overtraining is, in general, less critical than in the case of Boosted Decision Trees. 
The algorithm is validated on one- and two-dimensional analytically tractable toy models. 
An implementation of the algorithm is available on GitHub~\cite{BIT-algo}, and an introduction to the interface is provided in~\ref{sec:python}.

\section*{Acknowledgements}

R. S. and L. L. want to thank Andrea Valassi, author of Refs.~\cite{Valassi:2019uhy,Valassi:2020deh}, for useful discussions in the early stages of this project. 

\paragraph{Funding:} The work of D. S. and L.L. was supported by the Austrian Science Fund (FWF) projects P33771 and P31578, respectively. 

\clearpage
\appendix
\section{Python implementation and interface}\label{sec:python}

An implementation of the BIT algorithm in both Python\,2 and Python\,3 is available on GitHub~\cite{BIT-algo},  including a demonstration using Jupyter Notebook. The python package NumPy \cite{harris2020array} is the only external dependency. 

\paragraph{Preparation} Import the Python modules, for Python\,2
\begin{python}
import numpy as np
from BoostedInformationTree import BoostedInformationTree
\end{python}
and for Python\,3:
\begin{python}
import numpy as np
from BoostedInformationTreeP3 import BoostedInformationTree
\end{python}

Three two-dimensional NumPy arrays are needed as input: the features, the event weights, and the corresponding weight derivatives as defined in Eq.~\ref{eq:training-data}. 
Training and test data sets are available for the toy models. The power-law model is loaded with:
\begin{python}
data_dir = 'data'
def load_data(name):
    return np.loadtxt('

features = load_data('training_features_power_law_model')
features = features.reshape(features.shape[0], -1)
weights = load_data('training_weights_power_law_model')
diffs = load_data('training_diff_weights_power_law_model')
\end{python}

\paragraph{Hyperparameters} The training configuration (hyperparameters) comprises the weak learner's maximum depth $D$ (\texttt{max\_depth}) and its minimum event count in the terminal nodes~(\texttt{min\_size}). 
The boosting constructs a number $B$~(\texttt{n\_tree}) of weak learners to approximate the score. The learning rate $\eta$ is specified by the parameter \texttt{learning\_rate}. 
Setting the boolean flag \texttt{calibrated} to \texttt{True} calibrates the output to the unit interval. If it is set to \texttt{False}, the approximation of the score function in Eq.~\ref{eq:algo-result} is returned. For the toy models, a reasonable starting point is:
\begin{python}
learning_rate = 0.02
n_trees       = 100
learning_rate = 0.2 
max_depth     = 2
min_size      = 50
calibrated    = False
\end{python}

\paragraph{Training \& Prediction} To start the training, the \texttt{boost} method of a suitably constructed \texttt{BoostedInformationTree} object is called:
\begin{python}
bit = BoostedInformationTree(
        training_features     = training_features,
        training_weights      = training_weights, 
        training_diff_weights = training_diff_weights, 
        learning_rate         = learning_rate, 
        n_trees               = n_trees,
        max_depth             = max_depth,
        min_size              = min_size,
        calibrated            = calibrated)

bit.boost()
\end{python}

The trained discriminator can be used to predict the score for a new feature vector or an array of feature vectors:

\begin{python}
test_feat = load_data('test_features_power_law_model')
test_feat = test_feat.reshape(test_feat.shape[0], -1)

bit.predict(test_feat[0]) # returns one prediction
bit.vectorized_predict(test_feat) # returns all predictions
\end{python}

\clearpage

\bibliographystyle{elsarticle-num}
\section*{References}
\bibliography{references}

\end{document}